\documentclass[pra,aps,floats,twocolumn,groupedaddress,floatfix,showpacs]{revtex4}
\usepackage{graphicx,amssymb,amsmath,relsize}
\def\real{\mathop{\rm Re}\nolimits}   
\def\imag{\mathop{\rm Im}\nolimits}   

\begin{document}
\title{Geometric phases in electric dipole searches with trapped spin-1/2 particles in general fields and measurement cells of arbitrary shape with smooth or rough walls}

\author{R.~Golub}
\affiliation{Physics Department, North Carolina State University, Raleigh, NC 27606, U. S. A.}
\author{A.~Steyerl}
\email{asteyerl@mail.uri.edu}
\affiliation{Department of Physics, University of Rhode Island, Kingston, RI 02881, U. S. A.}
\author{C.~Kaufman}
\affiliation{Department of Physics, University of Rhode Island, Kingston, RI 02881, U. S. A.}
\author{G.~M\"uller}
\affiliation{Department of Physics, University of Rhode Island, Kingston, RI 02881, U. S. A.}

\pacs{28.20.-v, 14.20.Dh, 21.10.Tg}

\begin{abstract}
The important role of geometric phases in searches for a permanent electric dipole moment of the neutron, using Ramsey separated oscillatory field nuclear magnetic resonance, was first noted by Commins [Am.~J.~Phys.~\textbf{59}, 1077 (1991)] and investigated in detail by Pendlebury $\textit{et al.}$ [Phys.~Rev.~A~\textbf{70}, 032102 (2004)]. Their analysis was based on the Bloch equations. In subsequent work using the spin density matrix Lamoreaux and Golub [Phys.~Rev.~A \textbf{71}, 032104 (2005)] showed the relation between the frequency shifts and the correlation functions of the fields seen by trapped particles in general fields (Redfield theory). More recently we presented a solution of the Schr\"odinger equation for spin-$1/2$ particles in circular cylindrical traps with smooth walls and exposed to arbitrary fields [Steyerl \textit{et al.}, Phys.~Rev.~A $\mathbf{89}$, 052129 (2014)]. Here we extend this work to show how the Redfield theory follows directly from the Schr\"odinger equation solution. This serves to highlight the conditions of validity of the Redfield theory, a subject of considerable discussion in the literature [e.~g., Nicholas $\textit{et. al.}$, Progr.~Nucl.~Magn.~Res.~Spectr. $\mathbf{57}$, 111 (2010)]. Our results can be applied where the Redfield result no longer holds, such as observation times on the order of or shorter than the correlation time and non-stochastic systems and thus we can illustrate the transient spin dynamics, i.e.~the gradual development of the shift with increasing time subsequent to the start of the free precession.

We consider systems with rough, diffuse reflecting walls, cylindrical trap geometry with arbitrary cross section, and field perturbations that do not, in the frame of the moving particles, average to zero in time. We show by direct, detailed, calculation the agreement of the results from the Schr\"odinger equation with the Redfield theory for the cases of a rectangular cell with specular walls and of a circular cell with diffuse reflecting walls.

\end{abstract}
\maketitle

\section{Introduction}\label{sec:I}

The experimental search for a permanent electric dipole moment (EDM) of the neutron or other particles is motivated by its potential to search for deviations from the Standard Model. The most sensitive experimental approach so far, for neutrons, uses ultracold neutrons (UCNs) trapped in a measurement cell together with co-magnetometer atoms like $^{199}$Hg. They are exposed to a weak static magnetic field $\mathbf{B}_{0}$ and a strong electric field $\mathbf{E}$ applied parallel or anti-parallel to $\mathbf{B}_{0}$. Employing the Ramsey method of separated oscillatory field magnetic resonance the experiment is designed to search for small changes of the Larmor frequency due to the presence of the electric field.

At the high level of sensitivity already achieved in previous EDM experiments \cite{BAK01,HAR01,PEN01} and expected to be improved in current EDM projects (e.~g., \cite{ALT01,FRE01,ITO01,KIM01,MAR01}) a significant source of potential systematic error appears to be geometric phases. Their role in EDM experiments was first noted by Commins \cite{COM01} and analyzed by Pendlebury $\textit{et al.}$ \cite{PEN01} on the basis of the Bloch equation for spin evolution in a time-dependent magnetic field. Subsequent work by Lamoreaux and Golub \cite{LAM01} and Barabanov $\textit{et al.}$ \cite{BAR01} used the Redfield approach \cite{RED01,ABR01,MCG01} based on the spin-density matrix and extended the scope of such studies to general NMR physics. The original magnetic field model with a constant vertical gradient \cite{PEN01} of the Larmor field was extended in Refs.~\cite{HAR02,CLA01,SWA01,PIG01} to include more general macroscopic fields and the microscopic field of a magnetic dipole.

In Ref.~\cite{STE01} we showed that the frequency shifts due to geometric phases can also be determined by directly solving the Schr\"odinger equation to second order in the perturbation. Solutions were presented for circular cylindrical traps with smooth walls and arbitrary macroscopic magnetic fields as well as for the microscopic field of a magnetic dipole.

The direct solution of the Schr\"odinger equation includes vertical spin oscillations and the non-stationary, transient spin dynamics, i.e., the gradual development of the shifts for an arbitrary number of wall reflections subsequent to the start of free spin precession in the Ramsey scheme.

In the present article we extend this approach to include diffuse reflectivity from rough walls and to vertical cylindrical measurement cells with arbitrary cross section. Their top and bottom are assumed to be perfectly flat. In Sec.~\ref{sec:II} we develop an analytic solution of the Schr\"odinger equation for perfectly diffuse reflection, in the $xy$ plane only, on the cylindrical walls, both for a constant gradient field in circular cylindrical geometry and for a more general field model where $\mathbf{B}_{0}$ and $\mathbf{E}$ are not exactly aligned.

Our work is the first to address directly the effects of diffuse wall reflections in different geometries. In Sec.~\ref{sec:III} we consider a generic cross section of the measurement cell and calculate the frequency shifts (that linear in $E$ and those quadratic in electric field or magnetic field) for cells in the form of a rectangle with arbitrary aspect ratio. Rectangular cells are foreseen for EDM project \cite{ITO01}. Our results are valid over the full range of in-plane particle velocity $v$ for given Larmor frequency $\omega_{0}$. We compare them with the simulations described in Sec.~\ref{sec:IV} and observe consistency between the two approaches. The non-adiabatic and adiabatic limits of frequency shift agree with the universal predictions of Refs.~\cite{MCG01,PIG01,GUI01}. In Sec.~\ref{sec:V} we establish the equivalence of our method with other approaches \cite{RED01,LAM01,SWA01}. The agreement is notable in view of the different sets of assumptions made by each.

The usual approach to these problems, the Redfield theory of frequency shifts and relaxation based on the von Neumann equation for the density matrix, makes use of a plethora of assumptions \cite{ABR01} and has resulted in a considerable literature concerning its range of validity, e.~g.~\cite{NIC01}. The density matrix represents the effects of both the quantum mechanics and statistical fluctuations of the system under study. In our approach, which involves first solving the Schr\"odinger equation, to second order in the perturbation, for the wave function for an arbitrary time varying field and then applying the statistical averaging to the result, there is a clear separation between the quantum and statistical effects. In fact our solution can be easily applied to cases where the Redfield result no longer holds, such as observation times shorter than the correlation time, and non-stochastic systems.

\begin{figure}[tb]
  \begin{center}
 \includegraphics[width=77mm]{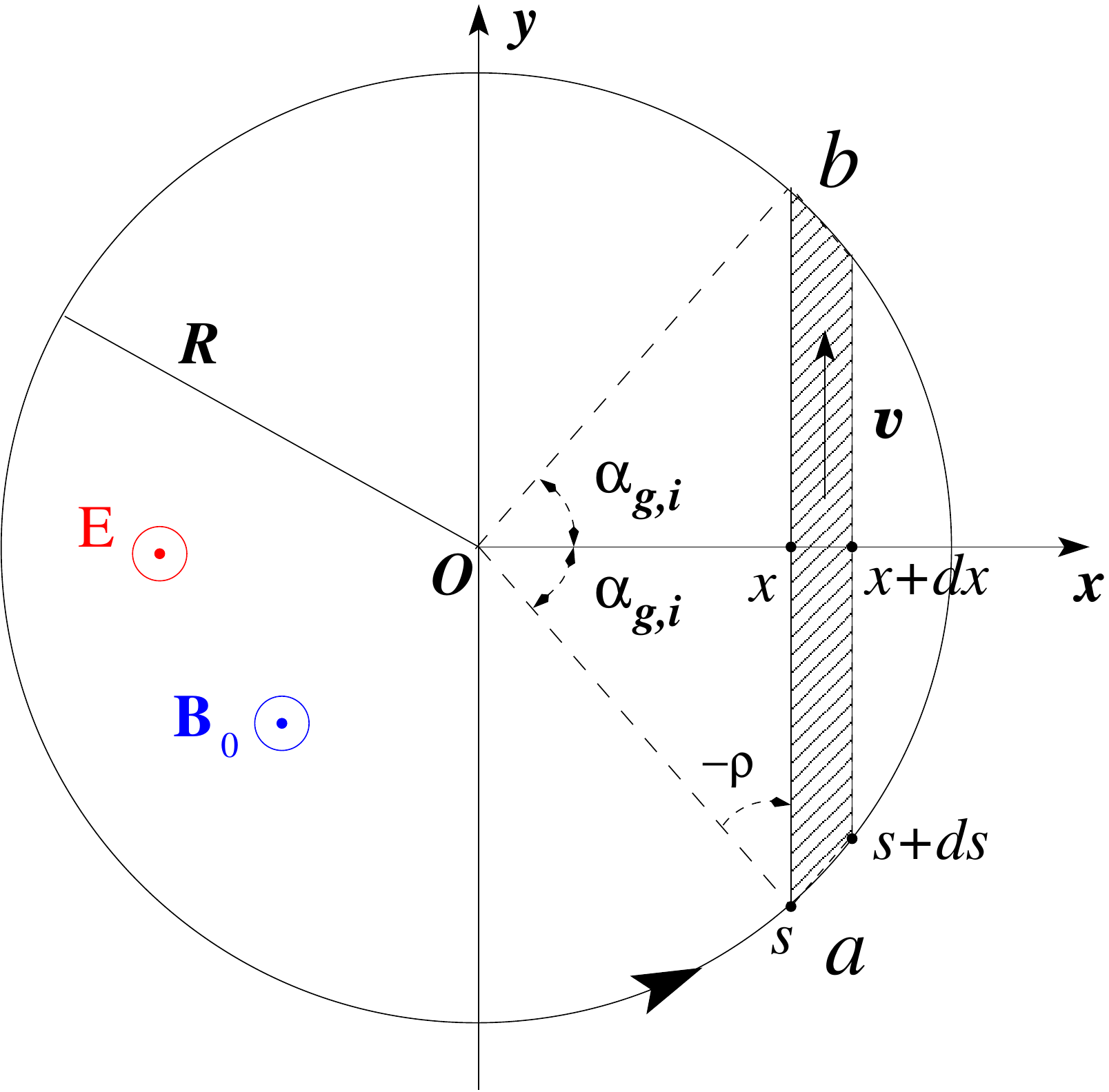}
\end{center}
\caption{(Color online) Trajectory of a particle moving ballistically in a circular cylindrical cell with vertical $z$-axis and radius $R$ between successive wall collisions at $a$ and $b$. A weak uniform magnetic Larmor field $\mathbf{B}_{0}$ and a strong uniform electric field $\mathbf{E}$ are applied along, or opposite to, the $z$ direction. The path segment $i$ shown can be characterized by the wall intersection angle $\rho$ or by the angle $\alpha_{g,i}$ where $2\alpha_{g,i}$ is the angle subtended by the chord as seen from the center $O$ of the cell.}
  \label{fig:one}
\end{figure}

\section{Frequency shift for a circular cylindrical trap with rough wall}\label{sec:II}

\subsection{Fundamentals}\label{sec:II A}

In the EDM experiments the cell walls are never perfect mirror surfaces, thus over the course of some $10^{1}$ to $10^{2}$ consecutive reflections randomization of trajectories is established for any initial distribution. The total number of wall reflections occurring during the measurement period in EDM experiments is much larger. 

In the present article we neglect scattering on gas molecules and consider perfectly diffuse surface scattering in the $xy$ plane perpendicular to the vertical symmetry axis of the cell (Fig.~\ref{fig:one}), thus it suffices to analyze the particle motion as projected onto the $xy$ plane \cite{PEN01}. The in-plane velocity $v$ is constant and the scattering probability distribution is $\propto\cos\rho$ (Lambert's law with in-plane scattering angle $\rho$), independent of incident angle $\rho_{inc}$ and, therefore, independent of the path history. Even a single UCN reflection will establish isotropy of flight direction. This is also a reasonable model for $^{199}$Hg co-magnetometer atoms which stick to the wall briefly before re-emission in a random direction.

Summarizing previous results \cite{STE01} we start from the Hamiltonian for a classical field,
\begin{equation}\label{1}
\mathcal{H} = -\mu\boldsymbol{\sigma}\cdot{\mathbf{B}} = \frac{\hbar}{2}\left[\begin{array}
[c]{cc}
\omega_{0} & \Sigma^{\ast}\\
\Sigma & -\omega_{0}
\end{array}\right]
\end{equation}
where $\sigma_{x}$, $\sigma_{y}$ and $\sigma_{z}$ are the Pauli matrices, $\omega_{0}=-2\mu B_{0}/\hbar$ is the Larmor frequency in the vertical static magnetic field $\mathbf{B}_{0}$ and $\mu$ is the magnetic moment of the particle. The measurement cell is also exposed to a uniform vertical electric field $\mathbf{E}$ which gives rise to the motional magnetic field $\mathbf{E}\times\mathbf{v}/c^{2}$.

In Eq.~(\ref{1}), $\Sigma(t)=\omega_{0}\left(B_{x}(t)+iB_{y}(t)\right)/B_{0}$ is the time-dependent perturbation due to the small horizontal magnetic field ($B_{x},B_{y}$) seen by the particle moving from point $a$ to point $b$ (Fig.~\ref{fig:one}) at constant $v$. In the case with small uniform vertical gradient $\partial B_{z}/\partial z$, first analyzed in \cite{PEN01}, the perturbing field is given as $B_{x}=-B_{0}(\eta\Omega+\zeta x/R)$, $B_{y}=-\zeta B_{0}y/R$, where we use the dimensionless parameters $\Omega=v/(R\omega_{0})$, $\zeta=(R/(2B_{0}))\,\partial B_{z}/\partial z$ and $\eta=R\omega_{0}E/(B_{0}c^{2})$.

For convenience we define a dimensionless time scale $\tau = \omega_{0}t$, reset the clock at the center of each chord and solve the Schr\"odinger equation for Hamiltonian (\ref{1}) in a coordinate system rotating about the vertical $z$ axis at the Larmor frequency $\omega_{0}$. For $\mathbf{B}_{0}$ pointing up, $\omega_{0}>0$ ($<0$) for $\mu<0$ as for neutrons (for $\mu>0$ as for $^{199}$Hg). Using a second-order perturbation approach we obtain the spin-up and spin-down components, $\alpha_{r}(\tau_{b},\tau_{a})$ and $\beta_{r}(\tau_{b},\tau_{a})$, of the wave function for a particle starting at point $a$ at time $\tau_{a}=-\delta_{i}$ with spin up and measured at $b$ at $\tau_{b}=+\delta_{i}$:
\begin{align}\label{2}
&\alpha_{r}(\delta_{i},-\delta_{i})=\textrm{e}^{-i\nu(\delta_{i},-\delta_{i})},\nonumber\\
&\beta_{r}(\delta_{i},-\delta_{i})=-\frac{i}{2}\Big(\Sigma_{\textrm{int}}(\delta_{i})-\Sigma_{\textrm{int}}(-\delta_{i}) \Big),
\end{align}
where $\delta_{i}=(\sin\alpha_{g,i})/\Omega$ and (see Eq.~(9) of \cite{STE01})
\begin{equation}\label{3}
\Sigma_{\textrm{int}}(t)=\int\! dt\,\textrm{e}^{-i\omega_{0}t}\Sigma(t).
\end{equation}

For circular cylindrical symmetry, vertical electric and magnetic fields $\mathbf{E}$ and $\mathbf{B}_{0}$ and uniform vertical gradient $\partial B_{z}/\partial z$, the functions $\nu$ and $\beta_{r}$ in (\ref{2}) are given by
\begin{align}\label{4}
&\nu(\delta_{i},-\delta_{i})=\frac{\zeta^{2}\Omega^{2}}{6}\Big\{\delta_{i}(3u^{2}_{1}+\delta_{i}^{2})\nonumber\\
&-3\sin\delta_{i}\Big[(u^{2}_{1}-\delta_{i}^{2})\cos\delta_{i} +2u_{1}\delta_{i}\sin\delta_{i}\Big] \Big\},\nonumber\\
&\beta_{r}(\delta_{i},-\delta_{i})=i\zeta\Omega\left(u_{1}\sin\delta_{i} -\delta_{i}\cos\delta_{i}\right)=\nonumber\\
&i\Omega\Big\{\eta\sin\delta_{i}+\zeta\Big[\left(1+\frac{\cos\alpha_{g,i}}{\Omega}\right)\sin\delta_{i}-\delta_{i}\cos\delta_{i} \Big] \Big\},
\end{align} 
as shown in \cite{STE01}, Eq.~(42). In (\ref{4}), $u_{1}=u+1$, $u=(\eta/\zeta)+(\cos{\alpha_{g,i}})/\Omega$. 

We write the wave function at the end of $n$ consecutive chords in the form
\begin{align}\label{5}
&\psi^{(n)}_{r}(\delta_{n})=\left[
\begin{array}
[c]{c}%
a^{(n)}(\delta_{n})\\
b^{(n)}(\delta_{n})
\end{array}
\right]=M^{(n)}_{r}(\delta_{n},\tau_{a,1})\psi^{(0)}\nonumber\\
&=M^{(n)}_{r}(\delta_{n},\tau_{a,1})\left[\begin{array}{c}\cos(\theta_{0}/2)
  \\
\sin(\theta_{0}/2)\textrm{e}^{i\Phi_{0}}
 \end{array}\right],
\end{align}
where the initial spinor $\psi^{(0)}$ points in an arbitrary direction given by angles $\theta_{0}$ (polar) and $\Phi_{0}$ (azimuthal). In the experiments we have $\theta_{0}\simeq\pi/2$. The recursion relation linking chains with $n-1$ and $n$ segments is (Eq.~(33) of \cite{STE01})
\begin{equation}\label{6}
M^{(n)}_{r}(\delta_{n},\tau_{a,1})=M_{T}(\delta_{n},-\delta_{n})R_{n}M^{(n-1)}_{r}(\delta_{n-1},\tau_{a,1}), 
\end{equation}   
where
\begin{equation}\label{7}
R_{n}=\left[
\begin{array}
[c]{cc}%
\textrm{e}^{i(\alpha_{g,n-1}+\alpha_{g,n})/2}  & 0 \\
0  & \textrm{e}^{-i(\alpha_{g,n-1}+\alpha_{g,n})/2}
\end{array}
\right]
\end{equation}
is the rotation matrix, around the $z$ axis, for angular change $\alpha_{g,n-1}+\alpha_{g,n}$ between the coordinate systems of segments $n-1$ and $n$. For each chord the $y$-axis is chosen to lie in the direction of motion.

In (\ref{5}) and (\ref{6}) we have left the starting position on chord $1$, given by $\tau_{a,1}$, arbitrary. The asymptotic frequency shift, for $n\gg 1$, does not depend on this position. Choosing point $a$ (Fig.~\ref{fig:one}) as the starting point of chord $1$, we set $\tau_{a,1}=-\delta_{1}$ and obtain for the initial transfer matrix $M^{(1)}_{r}(\delta_{1},\tau_{a,1})$ in (\ref{5})
\begin{equation}\label{8}
M^{(1)}_{r}(\delta_{1},-\delta_{1})=M_{T}(\delta_{1},-\delta_{1})
\end{equation} 
with
\begin{equation}\label{9}
M_{T}(\delta_{i},-\delta_{i})=T(-\delta_{i})M_{r}(\delta_{i},-\delta_{i})T(-\delta_{i}).
\end{equation}
The unitary evolution matrix along any chord $i$, in the rotating system, reads
\begin{equation}\label{10}
M_{r}(\delta_{i},-\delta_{i})=\left[
\begin{array}
[c]{cc}%
\alpha_{r}(\delta_{i},-\delta_{i})  & -\beta_{r}^{\ast}(\delta_{i},-\delta_{i}) \\
\beta_{r}(\delta_{i},-\delta_{i})  & \alpha_{r}^{\ast}(\delta_{i},-\delta_{i})
\end{array}
\right]
\end{equation}
and the transformation matrix from the laboratory to the rotating system at point $a$,
\begin{equation}\label{11}
T(-\delta_{i})=\left[
\begin{array}
[c]{cc}%
\textrm{e}^{-i\delta_{i}/2}  & 0 \\
0  & \textrm{e}^{i\delta_{i}/2}
\end{array}
\right],
\end{equation}
is the same as that for the reverse transition at point $b$.

When we perform the matrix multiplications in (\ref{6}) repeatedly for consecutive chords we obtain a sequence of matrices $M^{(n)}_{r}(\delta_{n},-\delta_{1})$ whose general form can be inferred by induction. Just as for a smooth wall the result is a unitary transfer matrix of the form
\begin{equation}\label{12}
M^{(n)}_{r}(\delta_{n},-\delta_{1})=\left[
\begin{array}
[c]{cc}%
g_{r}  & -h^{\ast}_{r} \\
h_{r}  & g^{\ast}_{r}
\end{array}
\right]
\end{equation}
for any $n\geq 2$. As shown in \cite{STE01}, Eq.~(\ref{35}), the cumulative phase advance at the end of chord $n$, relative to the rotating frame and averaged over initial azimuthal spinor angle $\Phi_{0}$ relative to initial flight direction, is given solely by the argument of $g_{r}$: 
\begin{equation}\label{13}
\langle\delta\varphi_{0\rightarrow n}\rangle=-2\arg g_{r}\simeq 2\imag{g^{\ast}_{r}},
\end{equation}
where the last form in (\ref{13}) holds since $g_{r}$ differs from unity only slightly.
 
For any $n\ge 2$ we obtain from recursion relation (\ref{6})
\begin{align}\label{15}
g^{\ast}_{r}&=\textrm{e}^{i\sum_{i=1}^{n}\nu_{i}}-\sum_{n_{0}=1}^{n-1}\!\beta^{\ast}_{r,n_{0}}\sum_{n_{1}=n_{0}+1}^{n}\!\!\!\beta_{r,n_{1}}\nonumber\\
&\times\, \textrm{e}^{i\left[\alpha_{g,n_{1}}-\delta_{n_{1}}+\alpha_{g,n_{0}}-\delta_{n_{0}}+2\sum_{k=n_{0}+1}^{n_{1}-1}(\alpha_{g,k}-\delta_{k})\right]}
\end{align}
where $\nu_{i}=\nu(\delta_{i},-\delta_{i})$ and $\beta_{r,i}=\beta_{r}(\delta_{i},-\delta_{i})$.

In general, reversing the order of two successive chords leads to a different phase advance. This property is reflected in the form of $g^{\ast}_{r}$. Expression (\ref{15}) is valid for any $n_{0}$, $n_{1}$ in the range $1\leq n_{0}<n_{1}\leq n$.

Using (\ref{15}) the overall phase shift (\ref{13}) reads
\begin{equation}\label{16}
\langle\delta\varphi_{0\rightarrow n}\rangle=2\sum_{i=1}^{n}\nu(\delta_{i},-\delta_{i})+Q
\end{equation}
with
\begin{align}\label{17}
Q=&-2\,\textrm{Im}\Big\{\!\sum_{n_{0}=1}^{n-1}\!\beta^{\ast}_{r,n_{0}}\sum_{n_{1}=n_{0}+1}^{n}\!\!\!\beta_{r,n_{1}}\\
&\times\, \textrm{e}^{i\left[\alpha_{g,n_{1}}-\delta_{n_{1}}+\alpha_{g,n_{0}}-\delta_{n_{0}}+2\sum_{k=n_{0}+1}^{n_{1}-1}(\alpha_{g,k}-\delta_{k})\right]}\Big\}.\nonumber
\end{align}

Dividing $\langle\delta\varphi_{0\rightarrow n}\rangle$ by the elapsed time $T_{0\rightarrow n}$ we obtain the average frequency shift $\langle\delta\omega \rangle_{0\rightarrow n}$. The averaging is subtle as described below in Sec.~\ref{sec:II B} after a closer look at the definition of angular advance $2\alpha_{g}$ around the cell per chord.

In Eq.~(\ref{7}) we have assumed that, going from chord $n-1$ to $n$, the net rotation angle to the new direction of travel can be written $\alpha_{g,n-1}+\alpha_{g,n}$. This is justified if, as in \cite{PEN01,STE01}, we define $\alpha_{g}$ as positive for passage on the right of the center $O$ (Fig.~\ref{fig:one}) (corresponding to counterclockwise rotation around the cell), as negative for passage on the left (for clockwise rotation), $\textit{and}$ if both chords pass $O$ on the same side (which is always true for specular reflection and circular geometry).

However, this definition for $\alpha_{g}$ (which we will later call $\alpha_{g,\textrm{old}}$) fails if in case of diffuse scattering the consecutive chords pass on opposite sides of $O$. If so the final direction of travel is opposite that given by $\alpha_{g,n-1}+\alpha_{g,n}$. This is evident for the special case $\alpha_{g,n}=-\alpha_{g,n-1}$ where the particle reverses its direction at the wall reflection point and thus has experienced a net rotation by angle $\pi$ rather than by $\alpha_{g,n,\textrm{old}}+\alpha_{g,n-1,\textrm{old}}=0$.

We avoid this inconsistency as follows: For passage on the right we define $\alpha_{g}$ as before: for circular geometry,
\begin{equation}\label{18}
\alpha_{g}=\rho+\pi/2,\,\,\textrm{  for }\rho<0.
\end{equation}  
For passage on the left ($\rho>0$) we choose, instead of the negative old value $\alpha_{g,\textrm{old}}=\rho-\pi/2$, continuation of (\ref{18}) to the range $\pi/2<\alpha_{g}<\pi$. Thus
\begin{equation}\label{19}
\alpha_{g}=\rho+\pi/2,\,\,\textrm{  also for }\rho>0.
\end{equation}
(\ref{19}) can also be written $\alpha_{g}=\alpha_{g,\textrm{old}}+\pi$. With this choice of a continuous function for $\alpha_{g}$, instead of the old function which changes abruptly from $+\pi/2$ to $-\pi/2$ at $\rho=0$, the net rotation angle can be expressed as $\alpha_{g,n-1}+\alpha_{g,n}$ also for passage on opposite sides. With this choice there is an offset by $2\pi$ if the next chord $n$ passes on the left, irrespective of the geometry of chord $n-1$. As a result of this offset, over the course of multiple consecutive chords a cumulative offset angle $2 m\pi$ may appear, with integer $m$, but this offset is irrelevant both for the analysis and the simulations presented below since these depend only on the sine and cosine of angles.

The choice of a continuous function for $\alpha_{g}$ throughout the entire range from $0$ to $\pi$, as in (\ref{18}), (\ref{19}), is also appropriate for the general cell geometries discussed below in section \ref{sec:III}.

\subsection{Two averaging procedures}\label{sec:II B}

As shown in \cite{PEN01} for circular cylindrical geometry and specular reflection, the frequency shift $X(\alpha_{g},\Omega)$ is averaged over all possible chord angles $\alpha_{g}$ using a weight factor $\sin^{2}\alpha_{g}$, thus
\begin{equation}\label{20}
\big\langle X(\alpha_{g},\Omega)\big\rangle_{1}=\frac{4}{\pi}\int_{0}^{\pi/2}\,d\alpha_{g}(\sin^{2}\alpha_{g})X(\alpha_{g},\Omega).
\end{equation}
We use subscript $1$ to distinguish this type of averaging from a second type introduced below in (\ref{23}).

First we show that the averaging scheme (\ref{20}) can also be derived differently and adapted to arbitrary cell geometry and rough surfaces satisfying Lambert's law. 

For circular geometry the take-off angle $\rho$ shown in Fig.~\ref{fig:one} is complementary to $\alpha_{g}$: $-\rho=\pi/2-\alpha_{g}$, implying $\cos\rho=\sin\alpha_{g}$. The quantity to be averaged is a frequency, i.e. the phase $\delta\varphi$ accumulated along the chord divided by the traversal time $L/v$, where $L$ is the chord length. Since $v=\textrm{const.}$, the frequency is determined by the phase per unit length, $\delta\varphi /L$. A longer chord represents a larger sample for averaging, thus the chord's contribution to the mean frequency is proportional to its length $L$. Including the Lambert factor $\cos\rho$ we obtain the weight factor $p_{1}(L,\rho)\propto L\cos\rho$. For circular geometry $L=2 R\cos\rho$, thus $p_{1}(\rho)\propto\cos^{2}\rho=\sin^{2}\alpha_{g}$ in agreement with the weight factor in (\ref{20}).

Changing from integration over $\alpha_{g}$ in (\ref{20}) to integration over $\rho$ we have
\begin{equation}\label{21}
\big\langle X(\rho,\Omega)\big\rangle_{1,\textrm{fw}}=\frac{1}{N_{1}} \int_{-\pi/2}^{\pi/2}d\rho\,(\cos^{2}\rho)X(\rho,\Omega),
\end{equation}
where the integral covers the full scattering range $-\pi/2<\rho<+\pi/2$. The normalization constant is $N_{1}=\pi/2$.

The measured phases are averages over both senses of particle circulation around the cell, forward (fw) and backward (bw). Eq.~(\ref{21}) represents the part for forward motion (in Fig.~\ref{fig:one}, from $a$ to $b$). The corresponding contribution for backward motion from $b$ to $a$ is
\begin{equation}\label{22}
\big\langle X(\rho,\Omega)\big\rangle_{1,\textrm{bw}}=\frac{1}{N_{1}} \int_{-\pi/2}^{\pi/2}d\rho\,(\cos^{2}\rho)X(\rho-\pi,-\Omega),
\end{equation}
where we replace $\rho$ by $\rho-\pi$ and $\Omega$ by $-\Omega$ to transform to the opposite direction of motion. The fw-bw even and odd contributions to the mean values are $\langle ... \rangle_{\textrm{even}}=(1/2)\big[\langle ... \rangle_{\textrm{fw}}+\langle ... \rangle_{\textrm{bw}} \big]$ and $\langle ... \rangle_{\textrm{odd}}=(1/2)\big[\langle ... \rangle_{\textrm{fw}}-\langle ... \rangle_{\textrm{bw}} \big]$, respectively.

Certain averages in (\ref{16}) do not involve relative quantities like the frequency (given by the phase per unit length) but properties of the chord as a whole, such as its length $L$ or angles $\alpha_{g}$ and $\rho$ or the traversal time $L/v=2\delta/\omega_{0}$. The contribution of each chord to these averages is independent of its length $L$, thus the weight factor changes to $p_{2}(\rho)\propto\cos\rho=\sin\alpha_{g}$. For these quantities the equivalence of averaging procedure (\ref{21}), (\ref{22}) is, for circular geometry,
\begin{align}\label{23}
&\big\langle Y(\rho,\Omega)\big\rangle_{2,\textrm{fw}}=\big\langle Y(\rho,\Omega)\big\rangle_{2,\textrm{even}}+\big\langle Y(\rho,\Omega)\big\rangle_{2,\textrm{odd}}\nonumber\\
&=\frac{1}{N_{2}} \int_{-\pi/2}^{\pi/2}d\rho\,(\cos\rho)\,Y(\rho,\Omega),\nonumber\\
&\big\langle Y(\rho,\Omega)\big\rangle_{2,\textrm{bw}}=\big\langle Y(\rho,\Omega)\big\rangle_{2,\textrm{even}}-\big\langle Y(\rho,\Omega)\big\rangle_{2,\textrm{odd}}\nonumber\\
&=\frac{1}{N_{2}} \int_{-\pi/2}^{\pi/2}d\rho\,(\cos\rho)\,Y(\rho-\pi,-\Omega),
\end{align} 
with $N_{2}=2$. For instance, $\langle\delta\rangle_{{2}}=\langle\delta\rangle_{{2,\textrm{even}}}=\pi/(4\Omega)$.

In summary, we use averaging $2$, given by (\ref{23}) for circular geometry, to average properties of the chord as a whole, such as $\delta$ or $\beta_{r}(\delta,-\delta)$ or $\textrm{e}^{2i(\alpha_{g}-\delta)}$ (as in Eq.~(\ref{28})). We average relative quantities such as frequency or $(\sin\delta)/\delta$ using method $1$ which is given by (\ref{21}), (\ref{22}) for circular geometry.

In Sec.~\ref{sec:III} we generalize these averaging prescriptions to a generic cylindrical cell geometry with arbitrary cross section.

\subsection{Determining the mean frequency shift}\label{sec:II C}
\subsubsection{For specular wall reflections}\label{sec:II C.1}

For specular reflection and circular geometry all consecutive chords are identical ($\alpha_{g,i}=\alpha_{g}$, $\delta_{i}=\delta$, $\nu_{i}=\nu$, $\beta_{r,i}=\beta_{r}$) and  Eqs.~(\ref{16}), (\ref{17}) for the net phase shift reduce to Eq.~(35) of \cite{STE01}:
\begin{align}\label{24}
&\langle\delta\varphi_{0\rightarrow n}\rangle=-2\arg g_{r}\nonumber \\
&=2n\nu(\delta,-\delta)-2 s_{\mu}^{2}\!\sum_{k=1}^{n-1}(n-k)\sin\big(2k(\alpha_{g}-\delta)\big)\nonumber \\
&\stackrel{n\gg 1}{\longrightarrow}2n\nu(\delta,-\delta)-n s_{\mu}^{2}\cot(\alpha_{g}-\delta)
\end{align}
with $s_{\mu}^{2}=|\beta_{r}(\delta,-\delta)|^{2}$. The corresponding asymptotic frequency shift (for $n\gg 1$) is Eq.~(37) of \cite{STE01}:
\begin{align}\label{25}
&\frac{\langle\delta\omega\rangle_{n\gg 1}}{\omega_{0}}=\left\langle\frac{\langle\delta\varphi_{0\rightarrow n}\rangle}{2 n\delta}\right\rangle_{1}=\\
&\left\langle\frac{\nu(\delta,-\delta)}{\delta}\right\rangle_{1}-\left\langle\frac{\beta^{\ast}_{r}(\delta,-\delta)\beta_{r}(\delta,-\delta)}{2\delta}\cot(\alpha_{g}-\delta)\right\rangle_{1},\nonumber 
\end{align}
where all mean values are fw-bw averages referring to frequency and, therefore, are of type 1 given by (\ref{21}), (\ref{22}). The results of \cite{STE01}, based on (\ref{25}), agree with those of \cite{PEN01,LAM01} 

\subsubsection{For perfectly diffuse wall reflections}\label{sec:II C.2}

From Eq.~(\ref{13}) the overall phase shift is determined by the $(1,1)$ element $g_{r}$ of transfer matrix $M^{(n)}_{r}(\delta_{n},-\delta_{1})$ and given by Eq.~(\ref{15}). We summarize the structure of $g_{r}$ as follows.

In (\ref{15}) the functions $\beta_{r}$ appear in the form of products $\beta_{r,n_{0}}\beta^{\ast}_{r,n_{1}}$ where $n_{0}$ and $n_{1}>n_{0}$ refer to different chords. The following analysis is significantly simplified by noting that for completely diffuse scattering the chords are statistically independent of one another. There is no memory of the previous path.

This is an essential point and argued as follows. For any trajectory emerging from a given surface element, such as at point $a$ in Fig.~\ref{fig:one}, the 2D Lambert law ensures isotropy of take-off direction $\rho$, independent of the incident angle. While this is true for any cell geometry, the circular geometry is unique in ensuring a second type of uniformity which is also required to render any two generations of chords (labeled by indices $i$ and $j$) independent of one another: a circular surface is illuminated uniformly by a diffuse point source positioned anywhere on the surface. Rays leaving point $a$ at angle $\rho$ within $d\rho$ intercept a surface strip of width $ds=L\,d\rho/\cos\rho$ where $L=2R\cos\rho$ is the chord length. Thus $ds=2R\,d\rho$ is independent of impact position. In consequence, at any step in any sequence of chords the wall is a uniform source of next generation chords. The orientation of the previous chord does not matter. We will see in Sec.~\ref{sec:III} that this is not true for geometries different from circular and in those cases we will have to resort to approximations. We obtain an exact solution only for circular geometry.

In the latter case, the average of a product of chord characteristics such as $\beta_{r}$, $\alpha_{g}$, $\delta$ for different chords $i$ and $j$, or of any functions $F_{i}$, $G_{j}$ thereof, can be factorized:
\begin{equation}\label{26}
\langle F_{i} G_{j}\rangle=\langle F_{i}\rangle \langle G_{j}\rangle.
\end{equation}
In the case at hand, where $n_{1}\ne n_{0}$, this applies to the product of functions $F_{n_{0}}=\beta^{\ast}_{r,n_{0}}\textrm{e}^{i(\alpha_{g,n_{0}}-\delta_{n_{0}})}$ and $G_{n_{1}}=\beta_{r,n_{1}}\textrm{e}^{i(\alpha_{g,n_{1}}-\delta_{n_{1}})}$ in (\ref{15}). Since all chords are equivalent we obtain the same averages for any $n_{0}$ and $n_{1}$:
\begin{align}\label{27}
&\langle \beta^{\ast}_{r,n_{0}}\textrm{e}^{i(\alpha_{g,n_{0}}-\delta_{n_{0}})}\rangle=\langle \beta^{\ast}_{r}\textrm{e}^{i(\alpha_{g}-\delta)}\rangle_{2},\nonumber\\
&\langle\beta_{r,n_{1}}\textrm{e}^{i(\alpha_{g,n_{1}}-\delta_{n_{1}})}\rangle=\langle \beta_{r}\textrm{e}^{i(\alpha_{g}-\delta)}\rangle_{2},
\end{align}
where we use averaging scheme $2$ since these functions refer to a chord as a whole, as defined before (\ref{23}). 

The same product rule applies to the terms in the sum over $k$ in (\ref{15}) where $n_{0}<k<n_{1}$. These terms appear as products, such as $\textrm{e}^{2i(\alpha_{g,n_{0}+1}-\delta_{n_{0}+1})}\textrm{e}^{2i(\alpha_{g,n_{0}+2}-\delta_{n_{0}+2})}\textrm{e}^{2 i(\alpha_{g,n_{0}+3}-\delta_{n_{0}+3})}$ with average $D^3$, where we define \begin{equation}\label{28}
D=\langle\textrm{e}^{2i(\alpha_{g}-\delta)}\rangle_2. 
\end{equation}

For circular geometry the explicit form of this mean value is, from (\ref{23}):
\begin{align}\label{29}
&D=\langle\textrm{e}^{2i(\alpha_{g}-\delta)}\rangle_{2}=(1/2)(D_{\textrm{fw}}+D_{\textrm{bw}}),\nonumber\\
&D_{\textrm{fw}}=\int_{0}^{\pi/2}d\alpha_{g}(\sin\alpha_{g})\,\textrm{e}^{2i(\alpha_{g}-\delta(\alpha_{g},\Omega))}\nonumber\\
&=\int_{-\pi/2}^{0}d\rho(\cos\rho)\,\textrm{e}^{2i(\alpha_{g}(\rho)-\delta(\rho,\Omega))},\nonumber\\
&D_{\textrm{bw}}=\int_{0}^{\pi/2}d\alpha_{g}(\sin\alpha_{g})\,\textrm{e}^{2i(-\alpha_{g}-\delta(-\alpha_{g},-\Omega))}\nonumber\\
&=\int_{-\pi/2}^{0}d\rho(\cos\rho)\,\textrm{e}^{2i(\alpha_{g}(\rho-\pi)-\delta(\rho-\pi,-\Omega))}, 
\end{align}
where $\alpha_{g}(\rho-\pi)=-\alpha_{g}(\rho)+\pi$ and $\delta(\rho-\pi,-\Omega)=+\delta(\rho,\Omega)$. In (\ref{29}) we have separated the terms for fw and bw motion and used the symmetry between ranges $\rho<0$ and $\rho>0$: $\alpha_{g}(-\rho)=-\alpha_{g}(\rho)+\pi$, $\delta(-\rho,\Omega)=\delta(\rho,\Omega)$. For circular geometry and also for the symmetric geometries analyzed in Sec.~\ref{sec:III B} we have $D=D_{\textrm{fw}}=D_{\textrm{bw}}$.

For brevity let $X=\alpha_{g}-\delta$, $X_{0}=\alpha_{g,n_{0}}-\delta_{n_{0}}$, $X_{1}=\alpha_{g,n_{1}}-\delta_{n_{1}}$, $\beta_{0}=\beta_{r,n_{0}}$, $\beta_{1}=\beta_{r,n_{1}}$. Expressed in these terms the second contribution (\ref{17}) to the overall phase (\ref{16}) is given by the imaginary part, $Q$, of
\begin{align}\label{30}
&S=-2\sum_{n_{0}=1}^{n-1}\beta^{\ast}_{0}\sum_{n_{1}=n_{0}+1}^{n}\beta_{1}\textrm{e}^{i[X_{0}+X_{1}+2\sum_{k=n_{0}+1}^{n_{1}-1}X_{k}]}\nonumber\\
&=-2\sum_{n_{0}=1}^{n-1}\beta^{\ast}_{0}\textrm{e}^{iX_{0}}\langle\beta_{1}\textrm{e}^{iX_{1}}\rangle_{2}\\
&\times\sum_{n_{1}=n_{0}+1}^{n} \langle\textrm{e}^{2iX_{n_{0}+1}}\rangle_{2} \times\langle\textrm{e}^{2iX_{n_{0}+2}}\rangle_{2}\times \textrm{...} \times\langle\textrm{e}^{2iX_{n_{1}-1}}\rangle_{2}.\nonumber
\end{align}

Each average in the last sum of terms is equal to $D$. Therefore this sum is geometric: 
\begin{equation}
1+D+D^{2}+\textrm{...}+D^{n-n_{0}-1}=\frac{1-D^{n-n_{0}}}{1-D}\nonumber
\end{equation}
and Eq.~(\ref{30}) becomes
\begin{align}\label{31}
&S=-2\langle\beta^{\ast}_{r}\textrm{e}^{iX}\rangle_{2}\langle\beta_{r}\textrm{e}^{iX}\rangle_{2} \sum_{n_{0}=1}^{n-1}\frac{1-D^{n-n_{0}}}{1-D}\nonumber\\
&=-2\left[n-\frac{1-D^{n}}{1-D} \right]\frac{\langle\beta^{\ast}_{r}\textrm{e}^{iX}\rangle_{2}\langle\beta_{r}\textrm{e}^{iX}\rangle_{2}}{1-D}
\end{align}
where we have used Eq.~(\ref{27}). The second term in the square brackets is negligible for $n\gg1$ since the magnitude of $D$ is always significantly less than 1. A similar negligible correction appears if the particle starts on chord $1$ at a point different from $a$ (Fig.~\ref{fig:one}). In the EDM experiments using UCNs \cite{BAK01,HAR01,PEN01} the starting point is given by the beginning of the period of free spin precession in the Ramsey scheme and the particle can be at any point of the initial chord at this time. 
 
We derive the ensemble averaged frequency shift $\langle\delta\omega\rangle$ by dividing each chord phase $\delta\varphi$ by the passage time $2\delta/\omega_{0}$. Combining (\ref{16}), (\ref{17}) and (\ref{31}) the forward-backward averaged frequency shift for perfectly diffuse $2D$ wall reflectivity thus becomes
\begin{align}\label{32}
&\frac{\langle\delta\omega\rangle}{\omega_{0}}=\left\langle\frac{\nu(\alpha_{g},\Omega)}{\delta(\alpha_{g},\Omega)}\right\rangle_{1,\textrm{even}}-\frac{1}{2}\imag{Q_{1}},\textrm{    with }Q_{1}=\\
&\frac{\big[\langle\beta_{r}\textrm{e}^{iX}\rangle_{2}\langle\beta_{r}^{\ast}\textrm{e}^{iX}/\delta\rangle_{1}\big]_{\textrm{even}}+\big[\langle\beta_{r}\textrm{e}^{iX}/\delta\rangle_{1}\langle\beta_{r}^{\ast}\textrm{e}^{iX}\rangle_{2}\big]_{\textrm{even}}}{{1-D}} .\nonumber
\end{align}
In (\ref{32}) we use the arithmetic mean value of the two contributions in the numerator of $Q_{1}$. To show that this is justified we employ the following relation between the two averaging procedures $1$ and $2$: For any function $F$ of chord variables we have
\begin{equation}\label{33}
\langle F/\delta\rangle_{1}=\langle F \rangle_{2}/\langle\delta\rangle_{2}.
\end{equation}
For circular cell geometry this follows directly from the averaging definitions (\ref{21}) - (\ref{23}). Since $\delta=(\cos\rho)/\Omega$, we have
\begin{equation}\label{34}
\langle\delta\rangle_{2}=(1/N_{2})\int_{-\pi/2}^{\pi/2}d\rho\,(\cos\rho)\frac{\cos\rho}{\Omega}=N_{1}/(\Omega N_{2})
\end{equation}
with normalization constants $N_{1}=\pi/2$, $N_{2}=2$, and
\begin{align}\label{35}
&\langle F/\delta\rangle_{1}= \frac{1}{N_{1}}\int_{-\pi/2}^{\pi/2}d\rho(\cos^{2}\rho)\frac{\Omega}{\cos\rho}\,F\\
&=\frac{\Omega}{N_{1}}\int_{-\pi/2}^{\pi/2}d\rho(\cos\rho)\,F=\frac{\Omega N_{2}}{N_{1}}\langle F \rangle_{2}=\langle F \rangle_{2}/\langle\delta\rangle_{2},\nonumber
\end{align}
both for fw and bw motion. This proves (\ref{33}). Thus, the two contributions in the numerator of $Q_{1}$ in (\ref{32}) are equal and it was justified to use their arithmetic mean value.

Relation (\ref{33}) also holds for the generalization of averaging procedures $1$ and $2$ to general cell geometry introduced below in (\ref{53}) - (\ref{56}) of Sec.~\ref{sec:III A.2}.  

Using (\ref{33}) the expression for the fw-bw averaged frequency shift (\ref{32}) becomes
\begin{align}\label{36}
&\frac{\langle\delta\omega\rangle}{\omega_{0}}=\frac{\langle\nu\rangle_{2,\textrm{even}}}{\langle\delta\rangle_{2}}-\frac{1}{\langle\delta\rangle_{2}}\imag{\Big\{\frac{\langle\langle\beta^{\ast}_{r}\textrm{e}^{iX}\rangle_{2}
\langle\beta_{r}\textrm{e}^{iX}\rangle_{2}\rangle_{\textrm{even}}}{1-D}\Big\}}\nonumber\\
&=\frac{\langle\nu\rangle_{2,\textrm{even}}}{\langle\delta\rangle_{2}}-\frac{1}{2\langle\delta\rangle_{2}}\,\times\\
&\imag{\Big\{\frac{\langle\beta^{\ast}_{r}\textrm{e}^{iX}\rangle_{2,\textrm{fw}}\langle\beta_{r}\textrm{e}^{iX}\rangle_{2,\textrm{fw}}+\langle\beta^{\ast}_{r}\textrm{e}^{iX}\rangle_{2,\textrm{bw}}\langle\beta_{r}\textrm{e}^{iX}\rangle_{2,\textrm{bw}}}{1-D} \Big\}}\nonumber.
\end{align}
Eq.~(\ref{36}) can also be understood as: mean frequency shift = mean net phase shift divided by mean net time.

As an example we give the explicit form of (\ref{36}) for the frequency shift $\propto\eta^{2}$:
\begin{align}\label{37}
&\frac{\langle\delta\omega\rangle_{\eta^{2}}}{\omega_{0}}=\frac{\langle\nu_{\eta^{2}}\rangle_{2}}{\langle\delta\rangle_{2}}-\frac{\eta^{2}\Omega^{2}}{2 N^{2}_{2}\langle\delta\rangle_{2}}\imag{\Big\{(1-D)^{-1}Q_{2}\Big\}},\\
&Q_{2}=\Big[\int_{-\pi/2}^{\pi/2}\,d\rho (\cos\rho) \sin[\delta(\rho,\Omega)]\,\textrm{e}^{i[\alpha_{g}(\rho)-\delta(\rho,\Omega)]} \Big]^{2}\nonumber\\
&+\Big[\int_{-\pi/2}^{\pi/2}\,d\rho (\cos\rho) \sin[\delta(\rho,\Omega)]\,\textrm{e}^{i[-\alpha_{g}(\rho)-\delta(\rho,\Omega)]} \Big]^{2}\nonumber.
\end{align}
We have used the functions $\nu(\delta,-\delta)$ and $\beta_{r}(\delta,-\delta)$ from (\ref{4}) and the subscript $\eta^{2}$ for the terms $\propto\eta^{2}$, with $\nu_{\eta^{2}}=(\eta^{2}\Omega^{2}/4)\left(2\delta - \sin(2\delta)\right)$ and $\langle\nu_{\eta^{2}}\rangle_{2,\textrm{even}}/\langle\delta\rangle_{2}=(\Omega^{2}/2)\big[1-\Omega J_{1}(2/\Omega)\big]$ where $J_{1}$ is a Bessel function. The factor $\sin\delta$ in the integrands of (\ref{37}) is the part of $\beta_{r}/(i\Omega)$ which is $\propto\eta$. The first (second) term in $Q_{2}$ describes forward (backward) motion along a pair of chords.

In Sec.~\ref{sec:V B}, Eq.~(\ref{104}), we will rederive Eq.~(\ref{37}) in a fundamentally different way.  

\begin{figure*}[tb]
  \begin{center} 
 \includegraphics[width=76mm]{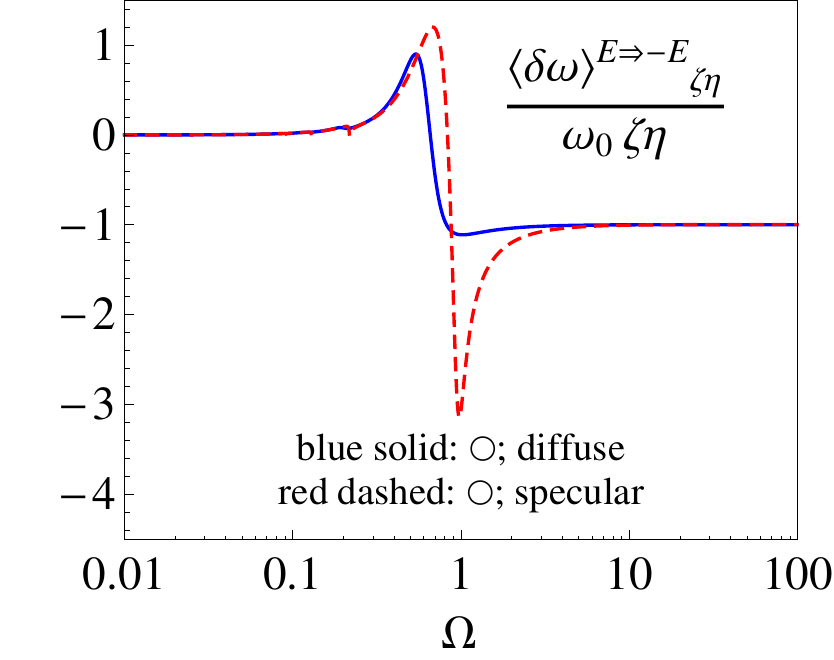}\hspace{0mm}%
 \includegraphics[width=76mm]{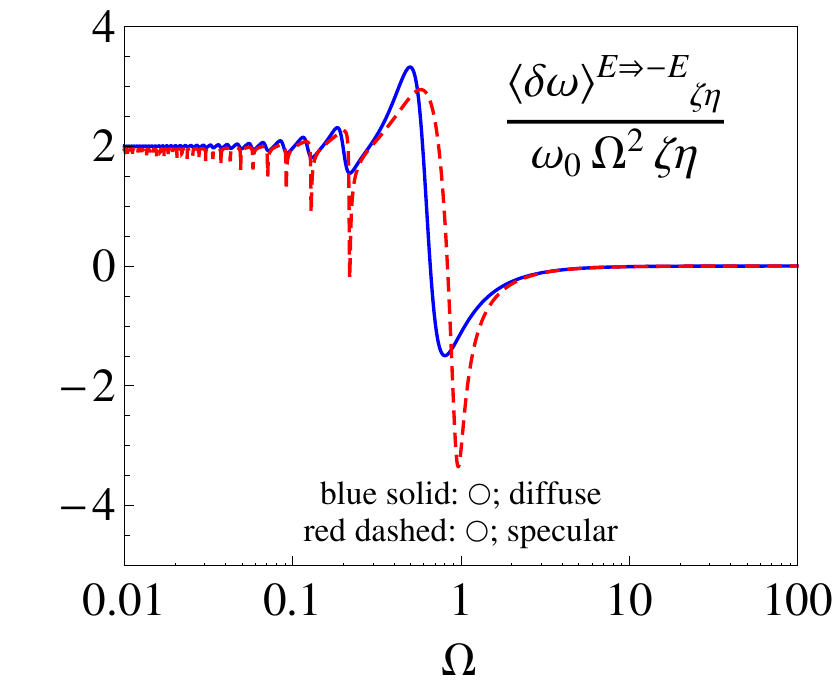}
  \includegraphics[width=76mm]{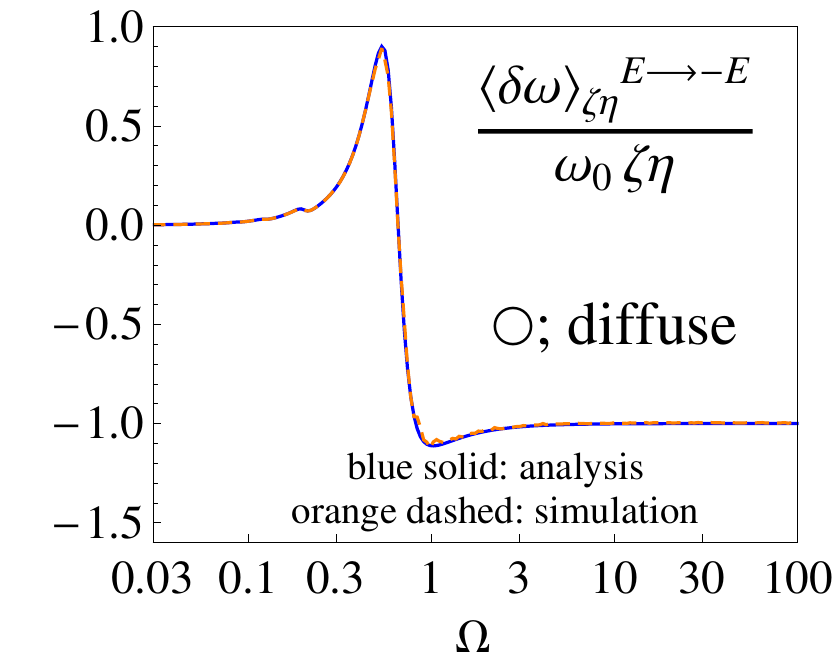}\hspace{0mm}%
 \includegraphics[width=76mm]{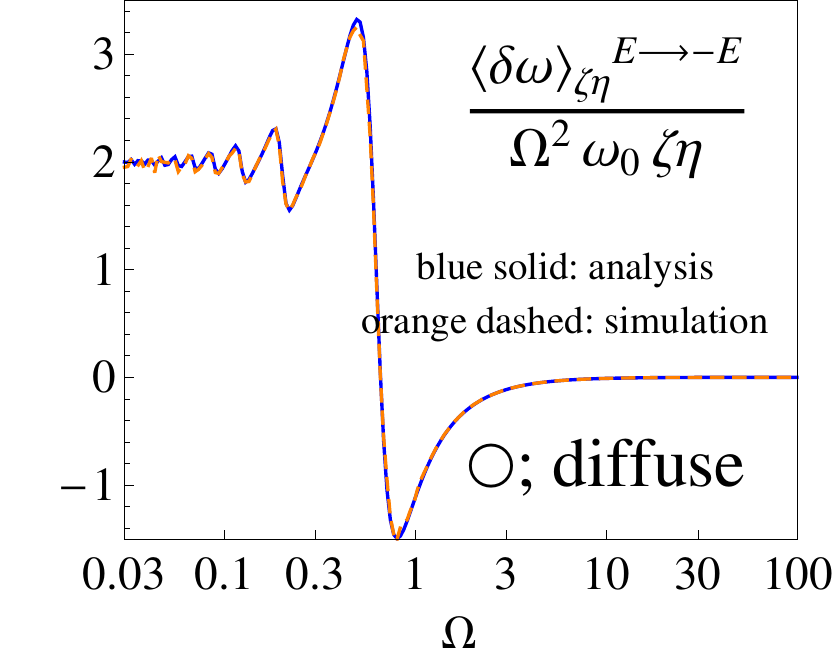}
\end{center}
\caption{(Color online) Upper panels: Normalized frequency shift for $E$-field reversal ($\propto E^{1}$) as a function of $\Omega=v/(R\omega_{0})$ for circular cell geometry. The data for perfectly diffuse 2D reflectivity (blue solid line) are compared to specular wall reflection (red dashed curve). On the right panel the shift is divided by $\Omega^{2}$ to show the asymptotic behavior in the adiabatic regime $\Omega\ll 1$. - Lower panels: Comparison of the calculation for diffuse reflection with simulations (orange dashed).}
  \label{fig:two}
\end{figure*}
\subsubsection{Comparison between specular and diffuse reflection}\label{sec:II C.3}

Equation (\ref{36}) is valid for perfectly diffuse 2D wall reflectivity. Comparing it with expression (\ref{25}) for specular reflectivity and noting identity (\ref{33}) we see that the first term on the right hand side (RHS) of (\ref{36}), which represents the single chord contribution, is the same as in (\ref{25}), thus independent of surface roughness.

The second term in (\ref{36}) ensures that the spinor remains unchanged at each wall reflection. In the limit of specular reflection it reduces to the second term in (\ref{25}) as follows: In this case $\alpha_{g}$ is constant along an orbit, thus $D=\textrm{e}^{2 i (\alpha_{g}-\delta)}$. The product of two averages in (\ref{36}) is replaced by the average of the single function $\beta^{\ast}_{r}\beta_{r}\,\textrm{e}^{2 i (\alpha_{g}-\delta)}$ of $\alpha_{g}$ over a random distribution of $\alpha_{g}$-values. Writing $\imag{\big[1/(1-\textrm{e}^{2 i (\alpha_{g}-\delta)})\big]}=\real{\big[\textrm{e}^{-i(\alpha_{g}-\delta)}\big]/[2 \sin(\alpha_{g}-\delta)]}=(1/2)\cot(\alpha_{g}-\delta)$, applying (\ref{33}) and averaging over the fw and bw directions we recover (\ref{25}) from (\ref{36}).

Prior to comparing numerical results for the frequency shifts we extend the constant gradient field, considered so far, to include the possibility of a slight misalignment between the vertical $\mathbf{E}$ field and the Larmor field $\mathbf{B}_{0}$. This type of imperfection of order $10^{-3}$ rad or more is hardly avoidable in the experimental situation.

\subsection{Extension to general gradient field}\label{sec:II D}

In Refs.~\cite{SWA01,PIG01,STE01} the uniform gradient field has been extended to more general macroscopic fields $\mathbf {B}=\nabla\chi$ derived from a power series expansion of a magnetic potential $\chi(x,y,z)$, including terms of higher than second order, e.g. $xy$, $x^{4} $, $x^{3}y$, etc.  Since the functions $\nu(\delta,-\delta)$ and $\beta_r(\delta,-\delta)$ can be determined analytically as in \cite{STE01}, we can use Eq.~(\ref{36}) to find analytic solutions also for 2D diffuse reflection.

We discuss briefly the effect of tilting the Larmor field slightly away from vertical by adding to $\mathbf{B}_{0}$ a small uniform horizontal component $\mathbf{B}_{\rho 0}=(B_{x0},B_{y0})$. Slight misalignment between the volume averaged Larmor field and the symmetry axis $z$ is inevitable in the experiments and has been taken into account in \cite{LAM01,AFA01} with the result that the precession about the tilted $z^{\prime}$ axis at frequency $\omega^{\prime}_{0}=\omega_{0}\sqrt{1+B^{2}_{\rho 0}/B^{2}_{0}}$ gives rise to the same frequency shift relative to $\omega^{\prime}_{0}$ as for a vertical Larmor field with magnitude given by the field magnitude $B_{L}=\sqrt{B^{2}_{0}+B^{2}_{\rho 0}}$.

\begin{figure*}[t]
  \begin{center} 
 \includegraphics[width=76mm]{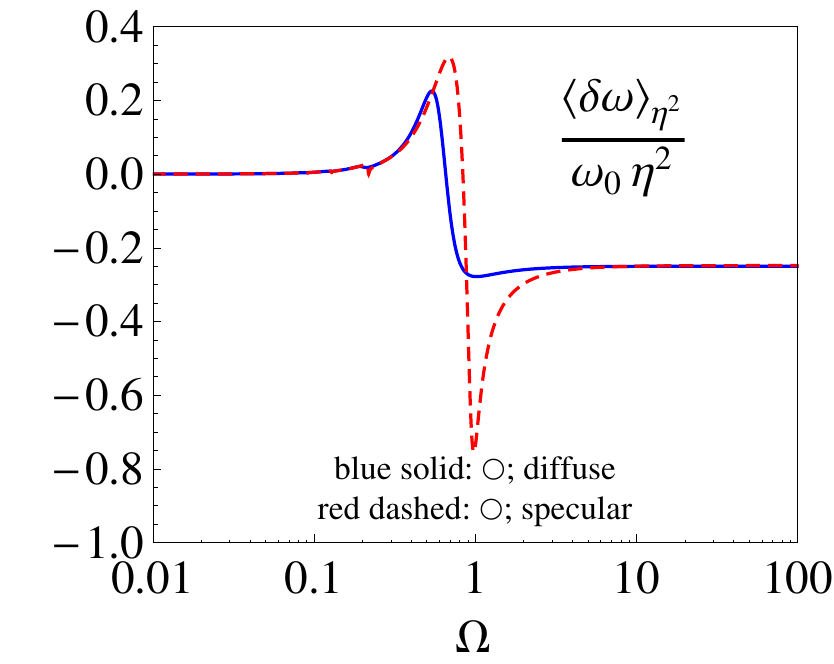}\hspace{0mm}%
 \includegraphics[width=76mm]{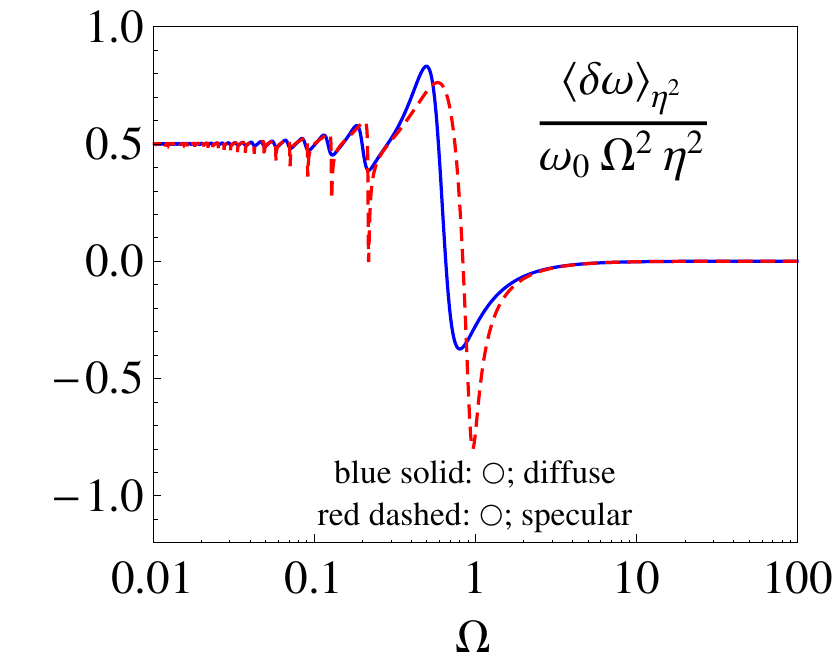}
\end{center}
\caption{(Color online) Comparison of normalized second-order motional frequency shift ($\propto E^{2}$) for diffuse reflectivity (shown by the blue solid curve) with that for specular reflection (red dashed line) as a function of $\Omega$, for circular cell geometry. On the right panel the shifts are divided by $\Omega^{2}$ to show the asymptotic behavior in the adiabatic regime $\Omega\ll 1$.}
  \label{fig:three}
\end{figure*}

\begin{figure*}[tb]
  \begin{center} 
 \includegraphics[width=76mm]{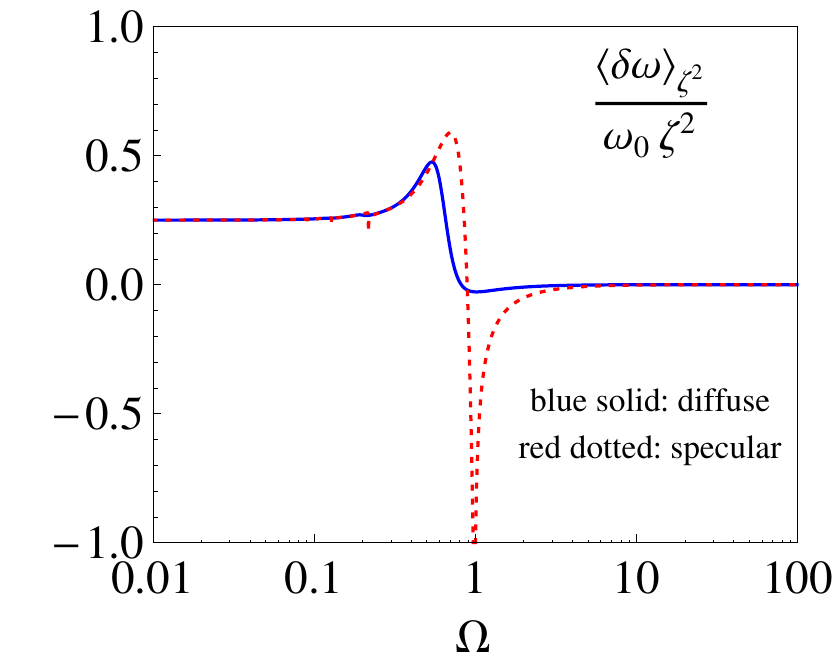}\hspace{0mm}%
 \includegraphics[width=76mm]{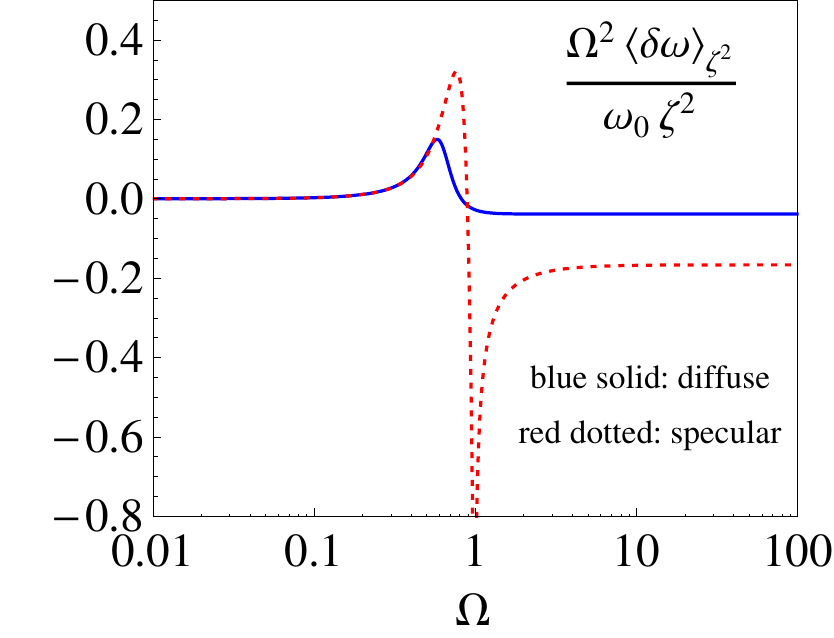}
\end{center}
\caption{(Color online) Comparison of normalized second-order magnetic frequency shift ($\propto B^{2}$) for diffuse reflectivity with that for specular reflection as a function of $\Omega$ for circular geometry. On the right panel the shifts are multiplied by $\Omega^{2}$ to show the asymptotic behavior in the non-adiabatic regime $\Omega\gg 1$.}
  \label{fig:four}
\end{figure*}

We can derive this result by exactly solving the Schr\"odinger equation for Hamiltonian (\ref{1}) in the $xyz$ laboratory frame. The two spinor components $\alpha(t)$, $\beta(t)$ are determined from \cite{STE01}
\begin{align}\label{38}
&\ddot{\alpha}(t)+\frac{1}{4}\omega^{\prime 2}_{0}\alpha(t)=0,\nonumber\\
&\beta(t)=\frac{B_{0}}{\omega_{0}(B_{x0}-i B_{y0})}\left(2 i \dot{\alpha}(t)-\omega_{0}\alpha(t) \right).
\end{align}
The solutions $\alpha(t)=\cos{(\omega^{\prime}t/2)}-i(\omega_{0}/\omega^{\prime}_{0})\sin{(\omega^{\prime}t/2)}$, $\beta(t)=i\left[(B_{x0}+i B_{y0})/B_{L}\right]\sin{(\omega^{\prime}t/2)}$, for initial spin up, describe uniform precession about the $z^{\prime}$ axis in the $xyz$ frame. In the limit of long time of free precession, $\omega^{\prime}_{0}t/(2\pi)\gg 1$, which is always realized in the experiments, the precession angles about the $z$ and about the $z^{\prime}$ axis become equal and are given by the field magnitude $B_{L}$.     

As a result, all results for frequency shifts relative to $\omega_{0}$, presented below for purely vertical volume average $\mathbf{B}_{0}$ of the field, also hold for a field with slightly tilted volume average, except that in this case the shift is measured relative to $\omega^{\prime}_{0}$. The redefinition of $\Omega$ in terms of $\omega^{\prime}_{0}$ rather than $\omega_{0}$, $\Omega\rightarrow v/(R\omega^{\prime}_{0})$, implies a negligible scale change in all plots of frequency shift vs.~$\Omega$. 

\subsection{Results for circular cell geometry}\label{sec:II E}

To separate the frequency shifts $\propto E^{1}$ (the $E$-odd shift for electric field reversal $E\rightarrow -E$, which we normalize in the form $\langle\delta\omega\rangle^{E\rightarrow -E}_{\zeta\eta}/(\omega_{0}\zeta\eta)$), the second-order motional shift $\propto E^{2}$ (denoted by $\langle\delta\omega\rangle_{\eta^{2}}/(\omega_{0}\eta^{2})$) and the second-order magnetic shift $\propto B^{2}$ (denoted by $\langle\delta\omega\rangle_{\zeta^{2}}/(\omega_{0}\zeta^{2})$) we select from Eq.~(\ref{25}) (for specular reflection), and from Eq.~(\ref{36}) (for diffuse reflection) the corresponding terms $\propto \zeta\eta$, $\propto \eta^{2}$, and, for $B^{2}$, those $\propto \zeta^{2}$.

Figures \ref{fig:two} to \ref{fig:four} show the results plotted against $\Omega$. We use parameters based on the EDM experiments at the ILL \cite{BAK01,HAR01,PEN01}: $E=10^{6}$ V/m, $B_{0}=1$ $\mu$T, $|\langle\partial B_{0}/\partial z \rangle|=1$ nT/m, averaged over a measuring cell with radius $R=0.235$ m and height $H=0.12$ m. For these values, $|\zeta|=0.00012$ and, for $^{199}$Hg co-magnetometer atoms, $|\eta|=0.0001$.

The integrations required in (\ref{36}) can be performed analytically, the results involving Bessel and Struve functions. However, most of the calculations were performed numerically.

In Figs.~\ref{fig:two} to \ref{fig:four} the specular data are the same as those of \cite{PEN01,STE01} except that in Fig.~2 of \cite{STE01} the dependence on finite $n$ had been shown explicitly and the $B^{2}$ shift had not been averaged over the fw-bw senses of circulation. The sharp resonances for specular reflection appear blurred since we used a slight imaginary offset of $\alpha_{g}$, away from the poles, in the numerical integration over $\alpha_{g}$.

The limiting behavior for $\Omega\ll 1$ is shown more clearly in the right panels of Figs.~\ref{fig:two} and \ref{fig:three} where the frequency shifts $\propto EB$ and $\propto E^{2}$ have been divided by $\Omega^{2}$. The lower panels of Fig.~\ref{fig:two} show the agreement, within numerical uncertainty, of $E$-odd shift values calculated from Eq.~(\ref{36}) with the simulations described below in Sec.~\ref{sec:IV}. We observed full agreement also for the second-order shifts $\propto E^{2}$ and $\propto B^{2}$. In Fig.~\ref{fig:four} the asymptotic form at large $\Omega$ is shown more clearly in the right panels, where the functions have been multiplied by $\Omega^{2}$.

Comparing the data for smooth walls and rough walls, the sharp resonances for specular reflection are largely washed out by roughness, as expected and previously shown by computer simulations in \cite{PEN01}.

Both the specular and diffuse reflectivity data in Fig.~\ref{fig:two} show agreement of the non-adiabatic limit $\Omega\rightarrow\infty$ for the $E$-odd shift with the universal Pignol limit [\cite{PIG01}, Eq.~(5)]
\begin{equation}\label{42} 
\left\langle\frac{\delta\omega}{\eta\omega_{0}}\right\rangle^{E\rightarrow -E}\stackrel{\Omega\gg 1}{\longrightarrow}\frac{2}{R B_{0}}\langle x B_{x}+yB_{y} \rangle.
\end{equation}
In circular geometry, the volume average is $\langle x B_{x}+yB_{y} \rangle=-\zeta R B_{0}/2$, thus $\left\langle\delta\omega/(\zeta\eta\omega_{0})\right\rangle^{E\rightarrow -E}\stackrel{\Omega\gg 1}{\longrightarrow}-1$.

\section{Extension to general cell shape}\label{sec:III}
\subsection{Generic cell cross section}\label{sec:III A}   

\begin{figure}[tb]
  \begin{center}
 \includegraphics[width=77mm]{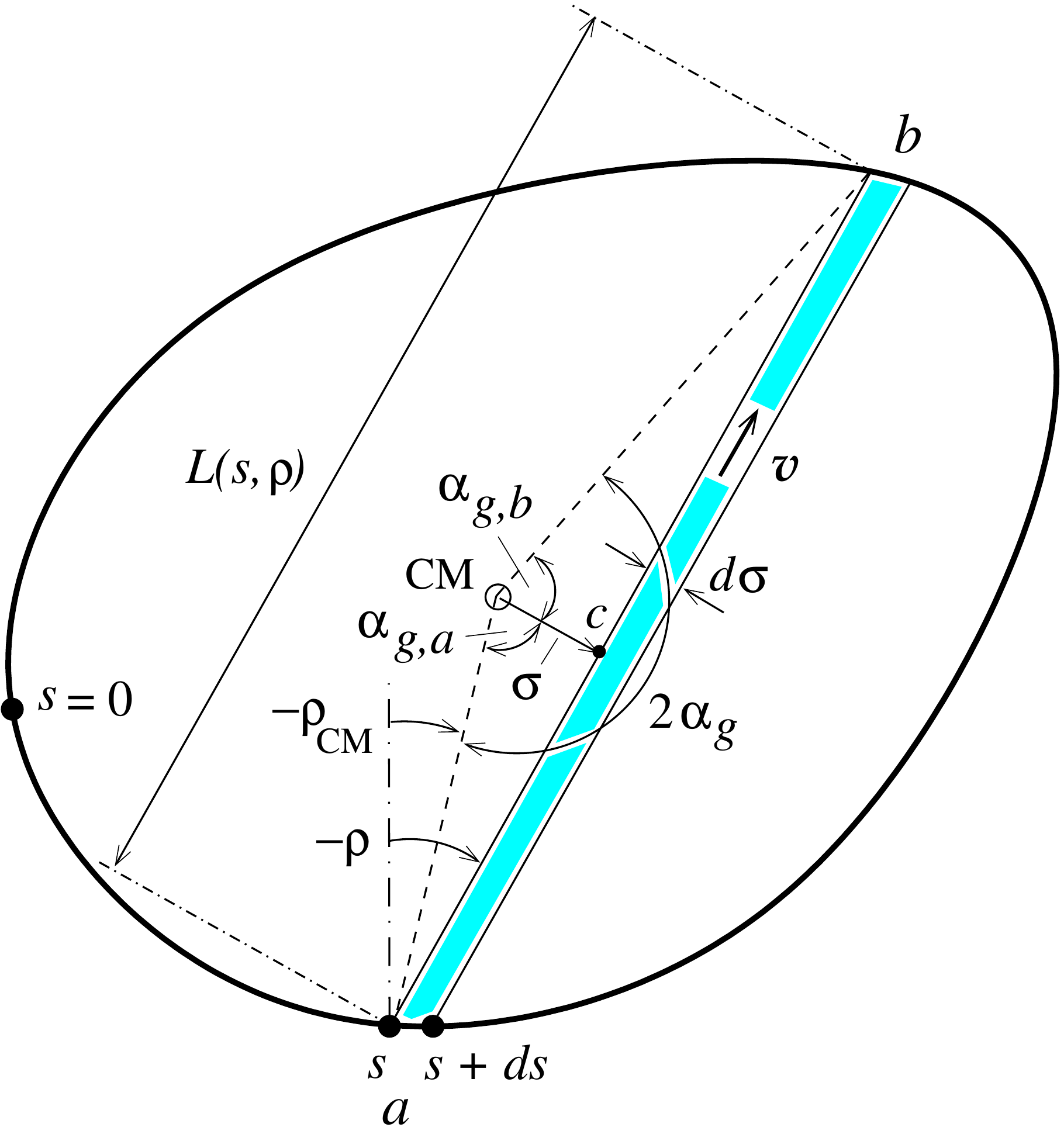}
\end{center}
\caption{(Color online) Measurement cell of generic shape subject to the restriction that any chord intersects the perimeter at two points only. The initial point $a$ of chord $ab$ is at location $s$ which is measured along the perimeter. As for circular geometry, $2\alpha_{g}$ designates the angle subtended by the chord as seen from the areal center of mass CM but the triangle $a\,\,b\,\,\textrm{CM}$ is, in general, asymmetric and the wall intersection angle $\rho$ is no longer directly determined by $\alpha_{g}$. $L$ and $\sigma$ are the length and ``lever arm'' of the chord. $L$ is divided into two sections of, generally, different length: from $a$ to the epicenter $c$ and from $c$ to $b$ with associated angles $\alpha_{g,a}$ and $\alpha_{g,b}$.}
  \label{fig:five}
\end{figure}

Figure \ref{fig:five} shows a cell of generic cross section. As in Fig.~\ref{fig:one} we show a trajectory of length $L$ extending from point $a$ to point $b$ where $a$ is at position $s$ measured along the perimeter from an arbitrary starting point $s=0$. The chord has the perpendicular distance $\sigma$ from the cross sectional center-of-mass point CM, thus $\sigma$ is the variable replacing the closest distance $x=R\cos\alpha_{g}$ for circular geometry. $\sigma$ is positive (negative) for trajectories passing the CM on the right (left).

As in Fig.~\ref{fig:one}, $2\alpha_{g}$ is the angle subtended by the chord $L$ as seen from the CM. It equals the angular advance around the cell when the particle travels from $a$ to $b$.

For circular geometry we had arbitrarily chosen the radius $R$ as a convenient reference length for the definition of $\Omega$, $\zeta$ and $\eta$: $\Omega=v/(R\omega_{0})$, $\zeta=R(\partial B_{z}/\partial z)/(2 B_{0})$ and $\eta=R\omega_{0}E/(B_{0}c^{2})$. For the rectangular cell shape analyzed below (and also for the equilateral triangle analyzed in \cite{STE02}) it is convenient to choose one half of the average side length as the unit of length. Thus, $\sigma$ and $L$ will denote normalized lengths which, for circular geometry, are given as $\sigma=\cos\alpha_{g}=-\sin\rho$ and $L=2\sin\alpha_{g}=2\cos\rho$. 

As before we measure the wall intersection angle $\rho$ from the surface normal and define it as positive for the counterclockwise direction (which corresponds to clockwise circulation of the particle around the cell). The trajectory at angle $\rho_{\textrm{CM}}$, which passes through the center point, divides the region on the right of CM from that on the left. We transform from forward to backward direction of motion by replacing $+\Omega$ by $-\Omega$ and $\rho$ by $\rho-\pi$. This operation reverses the direction of travel from $a$ to $b$ (in Figs.~\ref{fig:one} and \ref{fig:five}) to that from $b$ to $a$, retaining the sign of $\sigma$ and $\delta$, but reversing the sign of $\alpha_{g}$ (and adding an offset angle $\pi$, see (\ref{19})). The elapsed time is positive for either direction of motion. These symmetries are the same as for circular geometry.

\subsubsection{The role of chord asymmetry}\label{sec:III A.1}   

As viewed along a chord, a constant gradient field centered at CM is symmetric about the epicenter $c$, the point closest to CM. Thus, measuring time $\tau$ from point $c$ we have $B_{y}/B_{0}=-\zeta\Omega\tau$ and $B_{x}/B_{0}=-\zeta\sigma=$ const.

In general, the times at the chord endpoints, $\tau_{b}=\delta_{b}$ ($>0$ for the chord shown) and $\tau_{a}=-\delta_{a}$ ($<0$), will not add to zero. As for the circular case we define the average as $\delta=(\delta_{b}+\delta_{a})/2$. The imbalance is $\delta_{ba}=(\delta_{b}-\delta_{a})/2$. We also distinguish between two angles subtended from the center: $\alpha_{g,a}$ for the chord segment from $a$ to $c$, and $\alpha_{g,b}=2\alpha_{g}-\alpha_{g,a}$ for that from $c$ to $b$.

The imbalance $\delta_{ba}$ appears in the expressions for  $\nu(\delta_{b},-\delta_{a})$ and $\beta_{r}(\delta_{b},-\delta_{a})$:
\begin{align}\label{46}
&\nu(\delta_{b},-\delta_{a})=\frac{\zeta^{2}\Omega^{2}}{6}\Big\{(3u^{2}_{1}+\delta^{2})\delta\nonumber\\
&-3\sin\delta\Big[(u^{2}_{1}-\delta^{2})\cos\delta +2u_{1}\delta\sin\delta\Big] \Big\}\nonumber\\
&+\frac{\zeta^{2}\Omega^{2}}{2}\left(1-\frac{\sin 2\delta}{2\delta} \right)\delta^{2}_{ba},
\end{align}
\begin{align}\label{47}
&\beta_{r}(\delta_{b},-\delta_{a})=i\Omega\textrm{e}^{-i\delta_{ba}}\Big\{\sin\delta \big[\zeta \left(1+\frac{\sigma}{\Omega}\right)+\eta\big]\nonumber\\
&-\zeta\delta\cos\delta+i\zeta\delta_{ba}\sin\delta \Big\},
\end{align}
with $u_{1}=1+(\eta/\zeta)+\sigma/\Omega$.

The corresponding change of the transfer matrix element $g^{\ast}_{r}$ is taken into account in Eqs.~(\ref{17}), (\ref{30}), (\ref{31}), (\ref{32}) and (\ref{36}) by the replacements
\begin{align}\label{50}
X_{0}=&X_{n_{0}}=\alpha_{g,n_{0}}-\delta_{n_{0}}\nonumber\\
&\longrightarrow Y_{0}=\alpha_{g,b,n_{0}}-\delta_{b,n_{0}}\rightarrow \alpha_{g,b}-\delta_{b};\nonumber\\
X_{1}=&X_{n_{1}}=\alpha_{g,n_{1}}-\delta_{n_{1}}\nonumber\\
&\longrightarrow Y_{1}=\alpha_{g,a,n_{1}}-\delta_{a,n_{1}}\rightarrow \alpha_{g,a}-\delta_{a}.
\end{align}

In the limit $n\gg 1$ Eq.~(\ref{31}) becomes
\begin{equation}\label{51}
S=-2n\frac{\langle\left(\beta^{\ast}_{r}\textrm{e}^{iY_{0}}\right)\left(\beta_{r}\textrm{e}^{iY_{1}}\right)\rangle_{3} }{1-D}
\end{equation}
and the fw-bw averaged frequency shift takes the form
\begin{equation}\label{52}
\frac{\langle\delta\omega\rangle}{\omega_{0}}=\frac{\langle\nu\rangle_{2,\textrm{even}}}{\langle\delta\rangle_{2}}-\frac{1}{\langle\delta\rangle_{2}}\imag{\Big\{\frac{\langle\left(\beta^{\ast}_{r}\textrm{e}^{iY_{0}}
\right)\left(\beta_{r}\textrm{e}^{iY_{1}}\right)\rangle_{3}}{1-D} \Big\}}.
\end{equation}
As in Eq.~(\ref{36}) for circular geometry, the first term in (\ref{52}) is the single chord contribution. The second term ensures that the spinor remains unchanged at wall reflections. The two terms are averaged using the adaptation of methods $1$ and $2$ to general cell geometry described in the next section. We will also describe method $3$ for averaging the product $FG$ in (\ref{52}), with $F=\beta^{\ast}_{r}\textrm{e}^{iY_{0}}$ and $G=\beta_{r}\textrm{e}^{iY_{1}}$. For circular geometry this product had been averaged as in (\ref{36}). Its extension to general cell geometry is denoted $\langle (F)(G)\rangle_{3}$.

\subsubsection{Averaging procedures for general 2D cell geometry}\label{sec:III A.2}   

For circular cell geometry the surface coordinate $s$ is irrelevant and does not appear in the averaging procedures, (\ref{21}), (\ref{22}) for method $1$ and (\ref{23}) for method $2$. In an extension to general geometry we make use of the fact that, at equilibrium, all wall reflection points $s$ and chord directions are equally probable, as for a 2D gas. Therefore, at any time the probability, per unit length and unit time, for a trajectory to emanate from surface position $s$ is the same and the take-off angle $\rho$ obeys Lambert's distribution law $(\cos\rho)d\rho$. Thus, averaging procedures (\ref{21}) - (\ref{23}) are generalized to become (giving only the fw-bw even part in each case):
\begin{align}\label{53}
&\big\langle(...)\big\rangle_{1,\textrm{even}}=\frac{1}{2 N_{1}}\int_{\textrm{perimeter}}ds\,\,\int_{-\pi/2}^{\pi/2}d\rho\,(\cos\rho)L(s,\rho)\nonumber\\
&\times\,\left[(...)_{s,\rho,\Omega}+(...)_{s,\rho-\pi,-\Omega}\right],
\end{align} 
\begin{align}\label{54}
&\big\langle(...)\big\rangle_{2,\textrm{even}}=\frac{1}{2 N_{2}}\int_{\textrm{perimeter}}ds\,\,\int_{-\pi/2}^{\pi/2}d\rho\,(\cos\rho)\nonumber\\
&\times\,\left[(...)_{s,\rho,\Omega}+(...)_{s,\rho-\pi,-\Omega}\right]
\end{align} 
with normalization constants
\begin{align}\label{55}
&N_{1}=\int_{\textrm{perimeter}}ds\,\,\int_{-\pi/2}^{\pi/2}d\rho\,(\cos\rho)L(s,\rho)=2\pi A,
\end{align}
\begin{align}\label{56}
&N_{2}=\int_{\textrm{perimeter}}ds\,\,\int_{-\pi/2}^{\pi/2}d\rho\,(\cos\rho)=2 P
\end{align}
where $A$ is the cell area and $P$ is its perimeter.

Expression $N_{1}=2\pi A$ holds since, for reversed order of integration in (\ref{55}), the integral $\int ds\,(\cos\rho)\,\,L(s,\rho)$ represents the total area $A$ for all parallel strips of width $d\sigma=ds\cos\rho$, extending from the boundary on the left to that on the right in Fig.~\ref{fig:five}. The factor $2\pi$ in (\ref{55}) is the result of integration over a uniform distribution of strip directions.

To average the product $\left(\beta^{\ast}_{r}\textrm{e}^{iY_{0}}
\right)\left(\beta_{r}\textrm{e}^{iY_{1}}\right)$ in Eq.~(\ref{52}) we have to take account of the fact that consecutive chords $i$ and $i+1$ are not independent of one another. The endpoint of chord $i$ (which is at wall position $s$) coincides with the starting position of chord $i+1$.

We can quantify this correlation as follows: Consider trajectories starting from a given point $s$ in random direction. Except for the singular case of circular geometry discussed in Sec.~\ref{sec:II C.2}, the next wall impact point will be unevenly distributed over the wall surface and this non-uniformity extends over a number of consecutive reflections. This number increases with higher cell asymmetry, as for a rectangle with larger aspect ratio.

As a result, the averaging method used in Eq.~(\ref{36}) will be exact only for circular geometry. This is borne out by comparison of results for non-circular cells, based on (\ref{36}), with the simulations described below in Sec.~\ref{sec:IV}. It turned out that a closer approximation was obtained as follows. We average the product of functions $F(s,\rho,\Omega)=\beta^{\ast}_{r}\textrm{e}^{iY_{0}}$ and $G(s,\rho,\Omega)=\beta_{r}\textrm{e}^{iY_{1}}$ adopting a third method where we treat all pairs of chords as if they were consecutive with a common wall collision point $s$. This model leads to
\begin{align}\label{57}
&\langle F(s,\rho,\Omega)G(s,\rho,\Omega)\rangle_{3}=\frac{1}{2 N_{3}}\int_{\textrm{perimeter}}ds\nonumber\\
&\times\, \Big\{\Big[\int_{-\pi/2}^{\pi/2}d\rho\,(\cos\rho)F(s,\rho,\Omega)\Big]_{\textrm{fw}} \nonumber\\
&\times\,\Big[\int_{-\pi/2}^{\pi/2}d\rho\,(\cos\rho)G(s,\rho,\Omega) \Big]_{\textrm{fw}}+[\textrm{fw}\rightarrow \textrm{bw}]\Big\}
\end{align} 
with $N_{3}=4 P$. In Eq.~(\ref{57}) we average over the fw-bw directions using the fw$\rightarrow$bw transformations $\Omega\rightarrow -\Omega$, $\rho\rightarrow\rho-\pi$, $\sigma\rightarrow \sigma$, $(\alpha_{g},\alpha_{g,a},\alpha_{g,b})\rightarrow(-\alpha_{g}+\pi,-\alpha_{g,b}+\pi,-\alpha_{g,a}+\pi)$, $(\delta,\delta_{a},\delta_{b})\rightarrow(\delta,\delta_{b},\delta_{a})$ as for a time-reversed sequence of chords. 

As always, we obtain the frequency shift by dividing the net phase by the elapsed time and obtain Eq.~(\ref{52}) via identity Eq.~(\ref{33}).

Averaging schemes (\ref{53}) - (\ref{56}) are adapted to rectangular cells in (\ref{B7}), (\ref{B8}). For circular geometry, averaging procedures $1$, $2$ from (\ref{53}), (\ref{54}) reduce to (\ref{21}) - (\ref{23});  $2\langle\left(\beta^{\ast}_{r}\textrm{e}^{iY_{0}}
\right)\left(\beta_{r}\textrm{e}^{iY_{1}}\right)\rangle_{3}$ reduces to $\langle\beta^{\ast}_{r}\textrm{e}^{iX}\rangle_{2,\textrm{fw}}\langle\beta_{r}\textrm{e}^{iX}\rangle_{2,\textrm{fw}}+\langle\beta^{\ast}_{r}\textrm{e}^{iX}\rangle_{2,\textrm{bw}}\langle\beta_{r}\textrm{e}^{iX}\rangle_{2,\textrm{bw}}$ and thus the frequency shift (\ref{52}) reduces to Eq.~(\ref{36}).  

\subsection{Rectangular measurement cell}\label{sec:III B}

\begin{figure}[tb]
  \begin{center}
 \includegraphics[width=77mm]{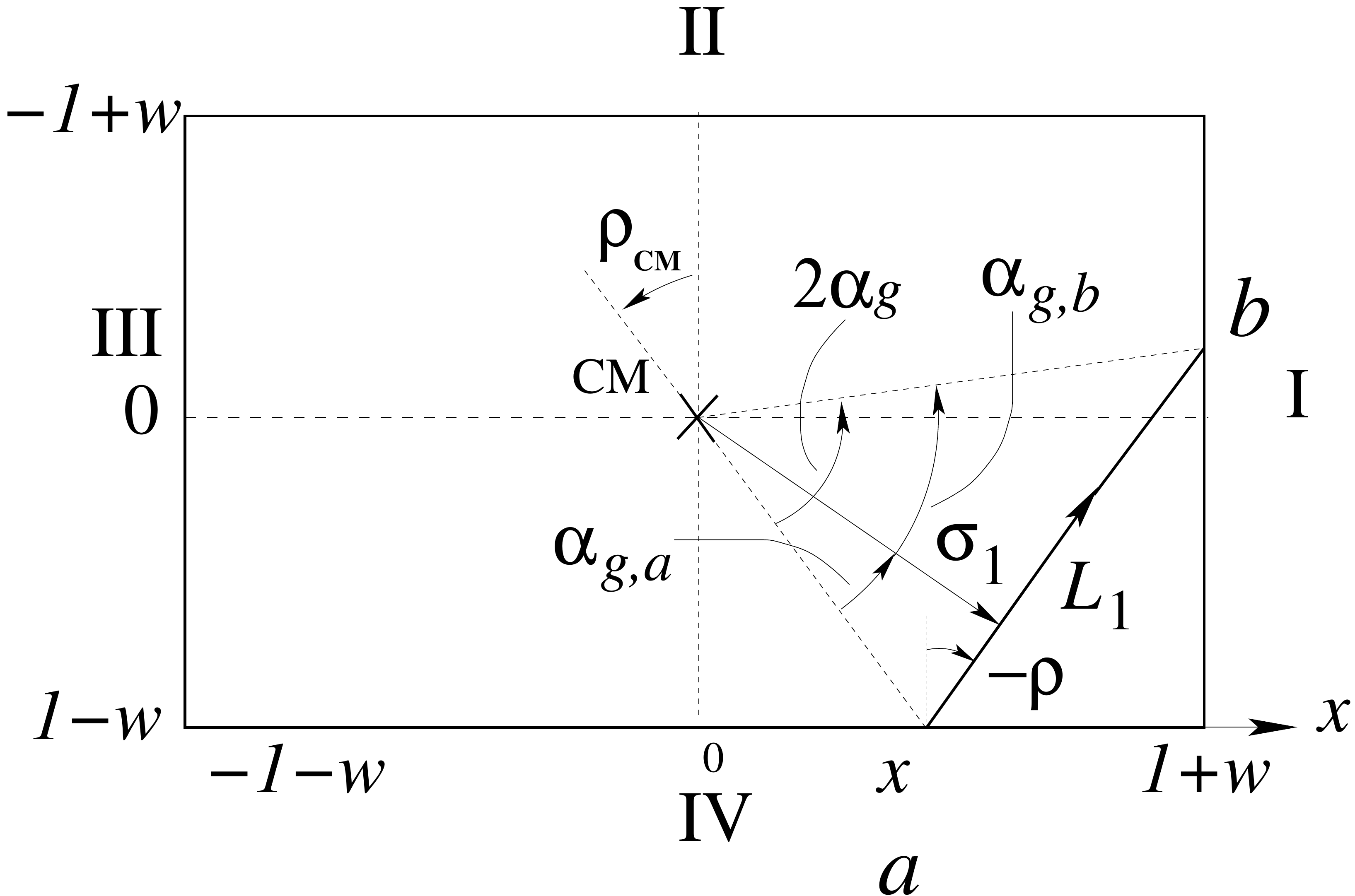}
\end{center}
\caption{ Rectangular measurement cell with side lengths $2(1+w)$ and $2(1-w)$, thus aspect ratio $(1+w)/(1-w)$. Among four groups of paths originating from side IV we show a path of group $1$ with endpoint on side I (see the text).}
  \label{fig:six}
\end{figure}
\begin{figure*}[h]
  \begin{center} 
 \includegraphics[width=76mm]{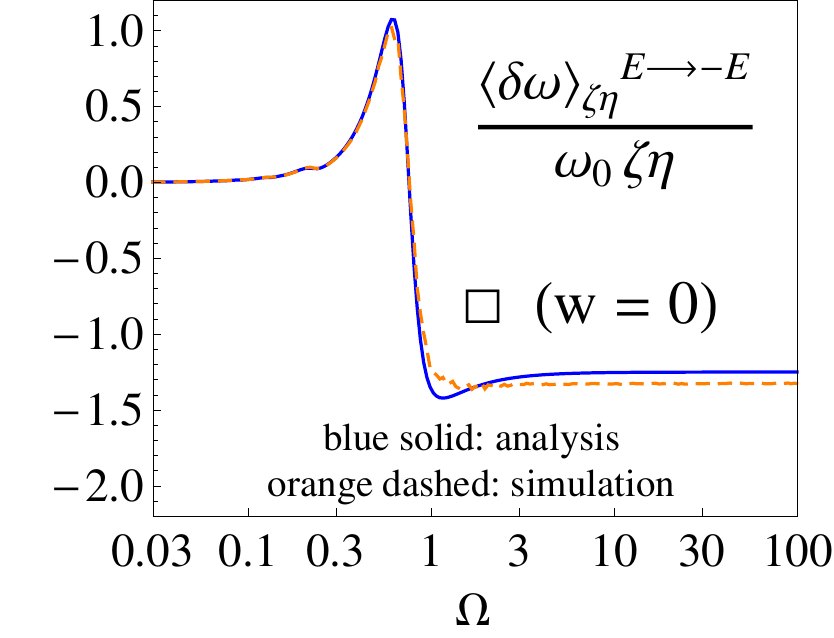}\hspace{0mm}%
  \includegraphics[width=76mm]{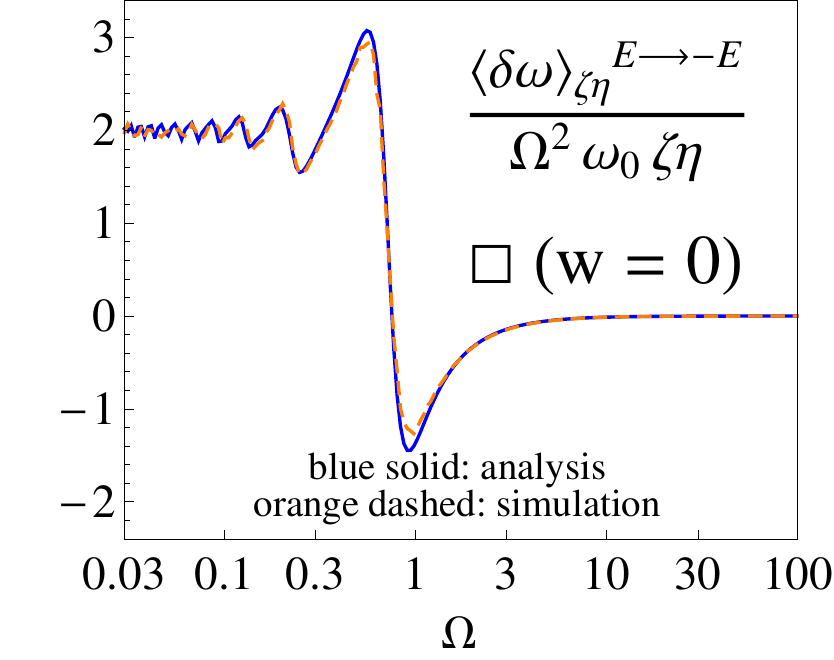}
   \includegraphics[width=76mm]{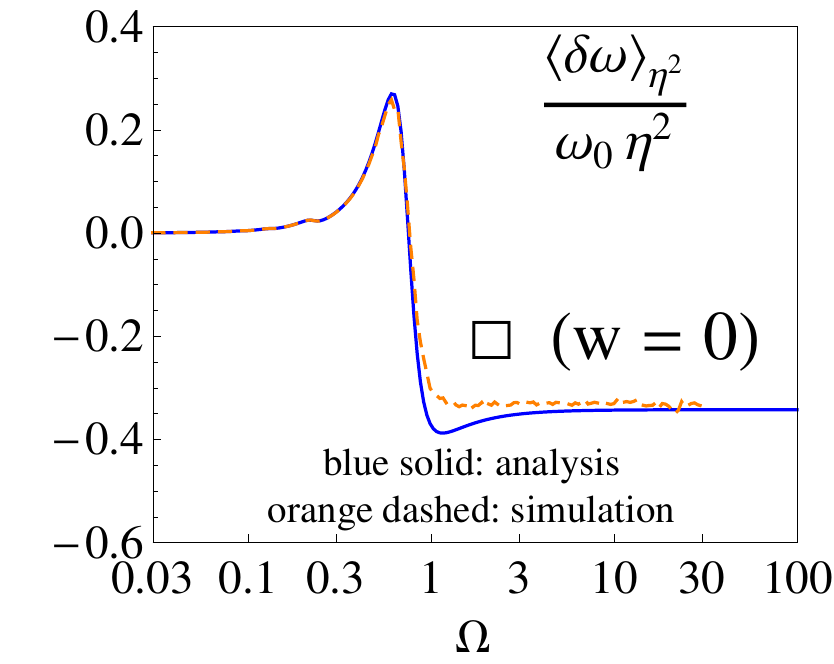}\hspace{0mm}%
  \includegraphics[width=76mm]{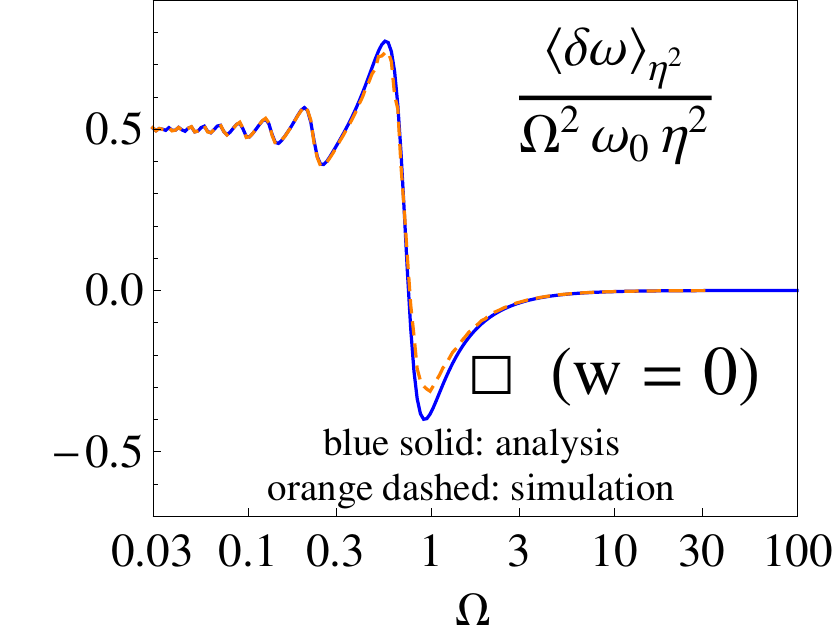}
   \includegraphics[width=76mm]{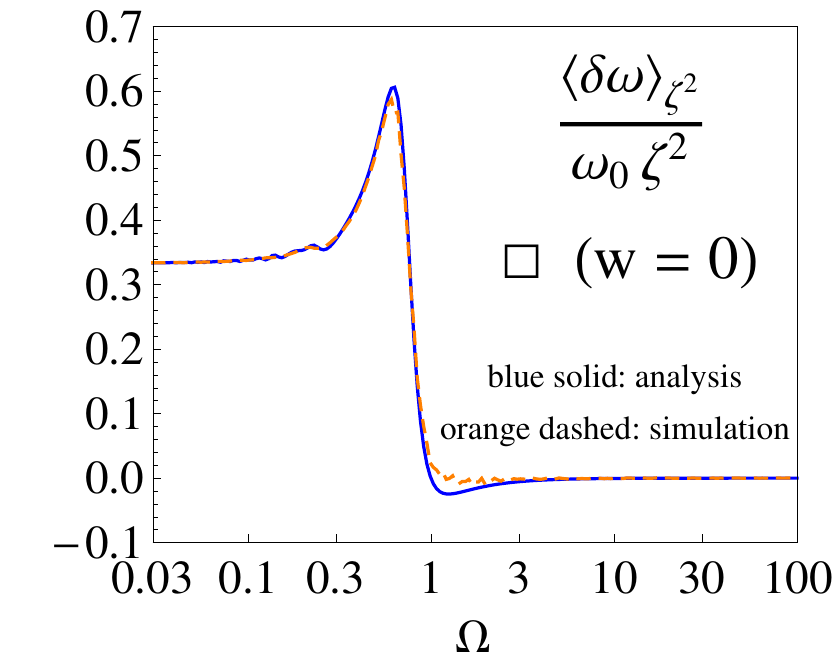}\hspace{0mm}%
  \includegraphics[width=76mm]{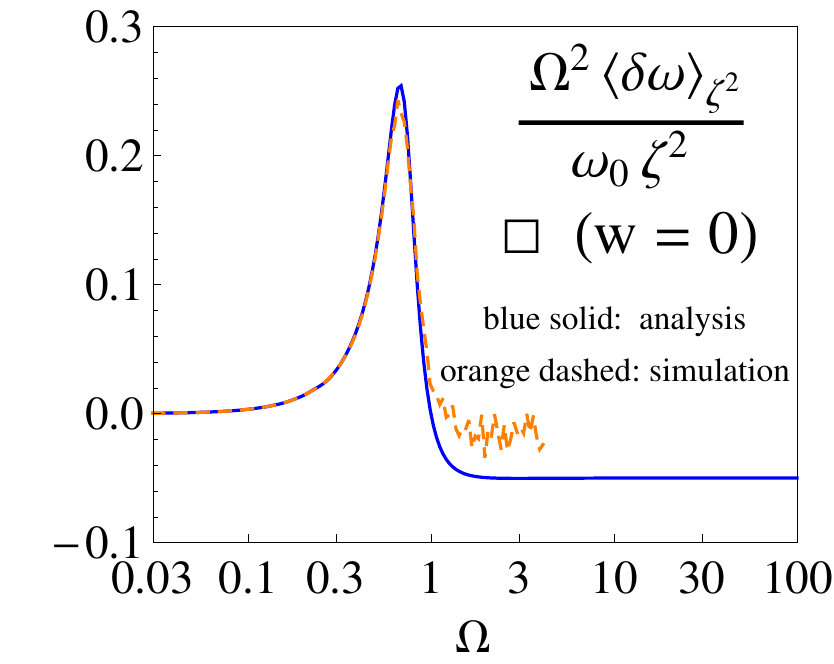}
\end{center}
\caption{(Color online) On the left: normalized frequency shifts from Eq.~(\ref{52}) for a square measurement cell ($w=0$) with rough walls as a function of $\Omega$. On top: Frequency shift for $E$-field reversal; center: $E^{2}$ shift; bottom: $B^{2}$ shift. On the right panels the asymptotic behavior for $\Omega\ll 1$ or $\Omega\gg 1$ is shown more clearly by division by $\Omega^{2}$ (top and center) or multiplication by $\Omega^{2}$ (bottom). For comparison we include the exact results from simulations.}
  \label{fig:seven}
\end{figure*}

\begin{figure*}[h]
  \begin{center} 
 \includegraphics[width=76mm]{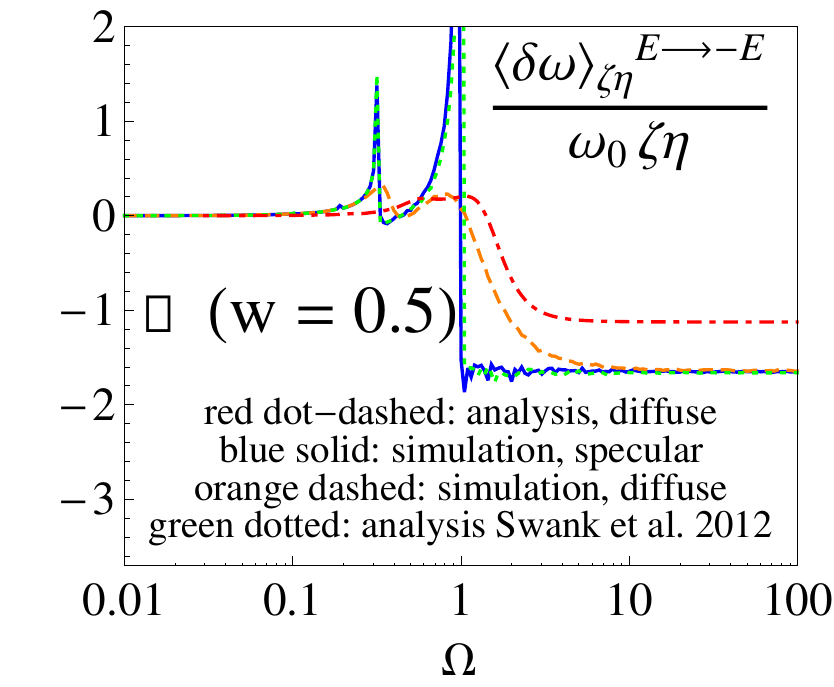}\hspace{0mm}%
  \includegraphics[width=76mm]{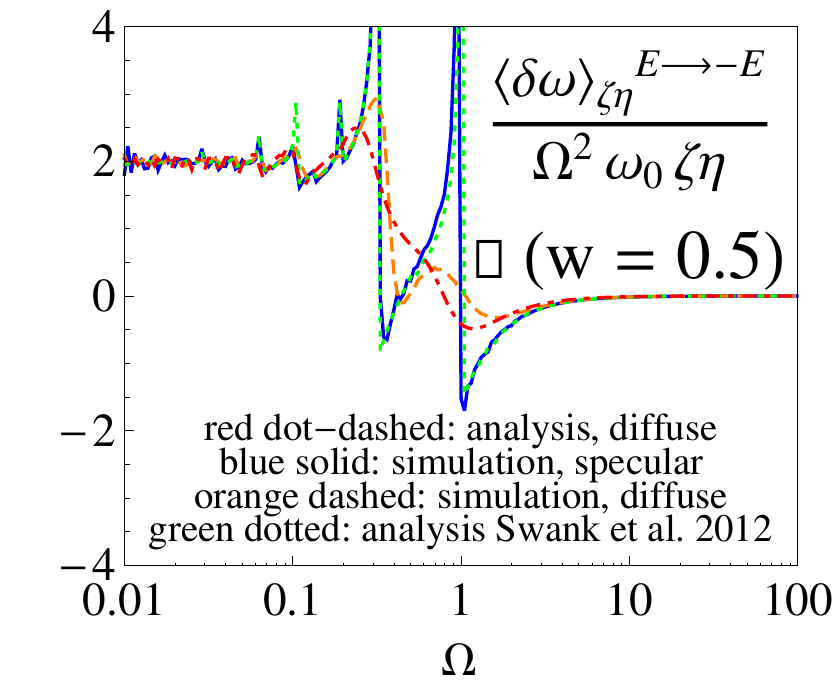}
   \includegraphics[width=76mm]{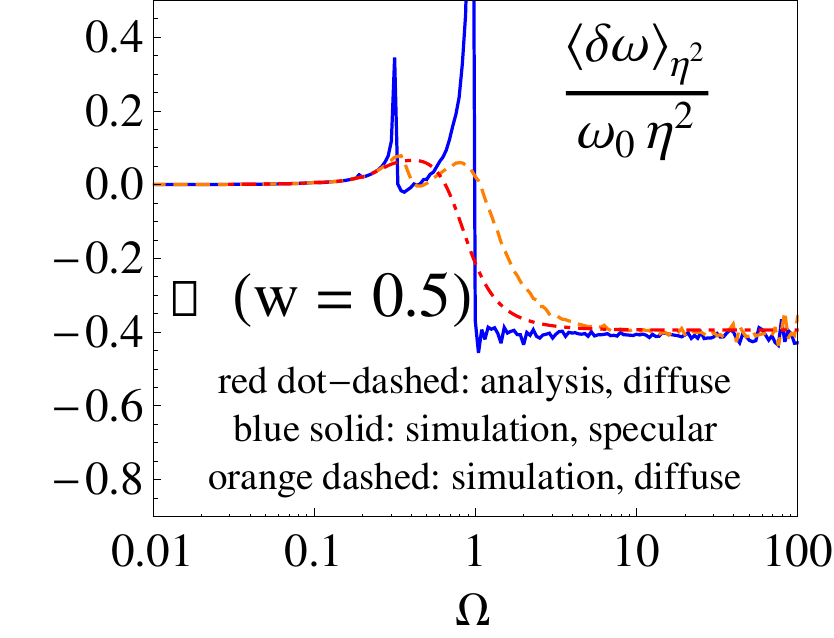}\hspace{0mm}%
  \includegraphics[width=76mm]{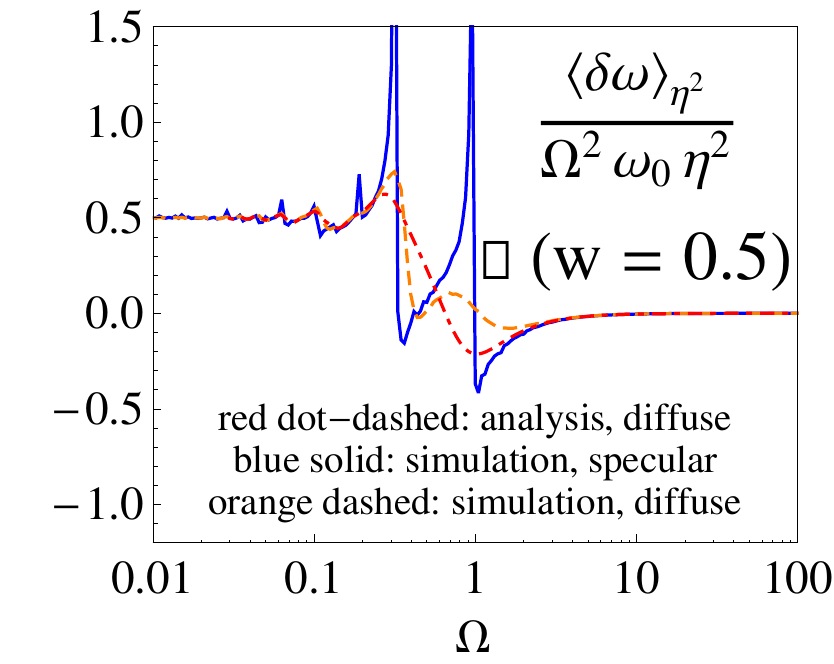}
   \includegraphics[width=76mm]{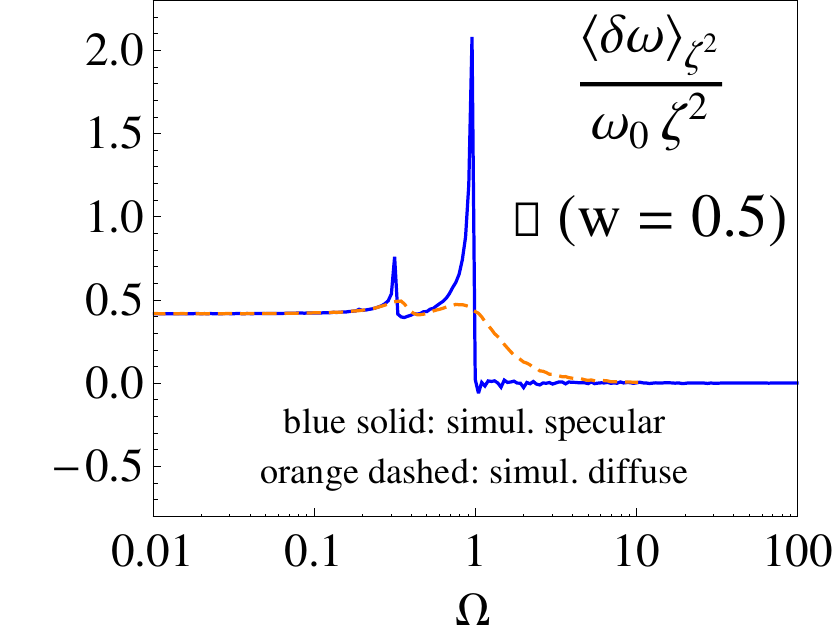}\hspace{0mm}%
  \includegraphics[width=76mm]{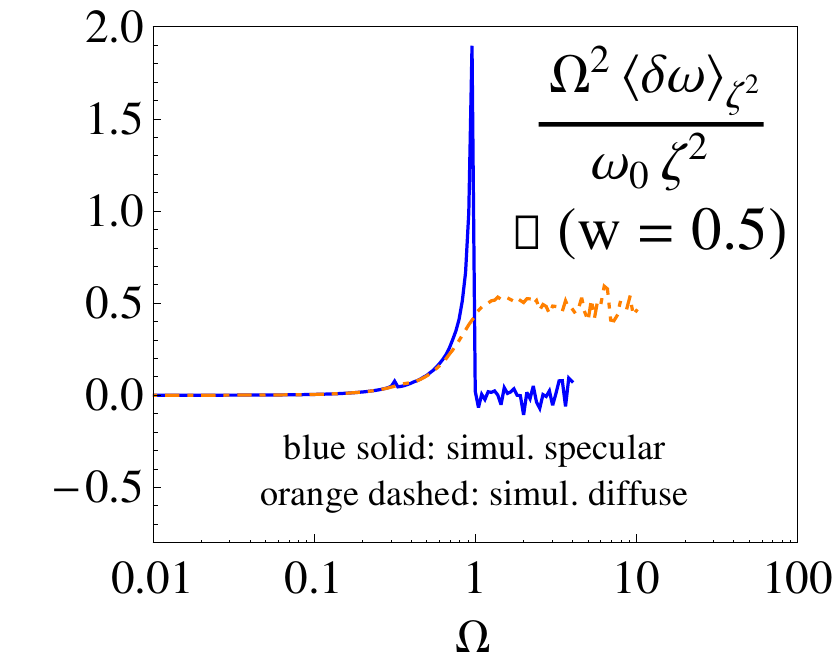}
\end{center}
\caption{(Color online) On the left: normalized frequency shifts for a rectangular measurement cell with aspect ratio 3 ($w=0.5$) as a function of $\Omega$. We compare the simulations and analysis for a smooth wall with those for perfect 2D roughness. On top: Frequency shift for $E$-field reversal; center: $E^{2}$ shift; bottom: $B^{2}$ shift. On the right panels the asymptotic behavior for $\Omega\ll 1$ or $\Omega\gg 1$ is shown by division by $\Omega^{2}$ (top and center) or multiplication by $\Omega^{2}$ (bottom). In the top panels we have included also the $E$-odd shift calculated for specular reflectivity on the basis of Swank $\textit{et al.}$ (2012) and described in Sec.~\ref{sec:V C}.}
  \label{fig:eight}
\end{figure*}

Figure \ref{fig:six} shows a rectangular cell with side lengths $2(1+w)$ and $2(1-w)$, where $w$ is in the range $0\leq w<1$ and determines the aspect ratio $(1+w)/(1-w)$. We use one eighth of the circumference as the unit of length. Referring to the analysis of this system in Appendix \ref{sec:B} we use Eqs.~(\ref{B1}) - (\ref{B9}) in approximation (\ref{52}) for the frequency shifts and in the simulations of Sec.~\ref{sec:IV} and plot the results in Figs.~\ref{fig:seven} and \ref{fig:eight} for $w=0$ (square cell) and $w=0.5$ (rectangle with aspect ratio 3). In Fig.~\ref{fig:seven} the values obtained from Eq.~(\ref{52}) for $w=0$ and diffuse wall scattering are seen to agree with the simulation data within a few percent.

In Fig.~\ref{fig:eight}, for $w=0.5$, we also compare simulations and analysis for specular walls with those for rough walls. While the adiabatic and non-adiabatic limits of the simulations are independent of roughness \cite{PEN01,LAM01,BAR01,PIG01,GUI01}, the resonances seen in the smooth-wall data are reduced by roughness. For the $E$-odd shift, the diffuse wall approximation (\ref{52}) deviates by $6\%$ ($33\%$) for $w=0$ ($w=0.5$) from the exact result (\ref{42}) of \cite{PIG01} which takes the form
\begin{equation}\label{59}
\Big\langle\frac{\delta\omega}{\omega_{0}\zeta\eta}\Big\rangle^{E\rightarrow -E}\stackrel{\Omega\gg 1}{\longrightarrow}-\frac{4}{3}(1+w^{2}),
\end{equation}
for arbitrary aspect ratio $(1+w)/(1-w)$.

For specular reflection we can compare the $E$-odd frequency shift data with a calculation based on an extension of the analysis of relaxation times by Swank $\textit{et al.}$ \cite{SWA01} to frequency shifts, which is described in Sec.~\ref{sec:V C}. In Fig.~\ref{fig:eight}, top panel, we observe agreement of this analysis with the simulations.

\subsection{Further common results for any cell geometry}\label{sec:III D}

In the adiabatic limit $\Omega\ll 1$ we can compare the second-order frequency shifts $\propto E^{2}$ and $\propto B^{2}$, for any cell geometry, with the universal expression (18) of \cite{GUI01}. As seen in Figs.~\ref{fig:three}, \ref{fig:four}, \ref{fig:seven} and \ref{fig:eight} we obtain, within statistics,
\begin{equation}\label{60}
\frac{\langle\delta\omega\rangle_{\eta^{2}}}{\omega_{0}}\stackrel{\Omega\ll 1}{\longrightarrow}\frac{1}{2}\eta^{2}\Omega^{2},\,\,
\frac{\langle\delta\omega\rangle_{\zeta^{2}}}{\omega_{0}}\stackrel{\Omega\ll 1}{\longrightarrow}\varsigma \zeta^{2}
\end{equation}
for all cell geometries and surface reflection laws investigated. For circular resp.~rectangular resp.~triangular geometry: $\varsigma=1/4$ resp.~$(1+w^{2})/3$ resp.~$1/6$, in agreement with Eq.~(18) of \cite{GUI01}.

We also recover the universal adiabatic limit
\begin{equation}\label{61}
\langle\delta\omega\rangle^{E\rightarrow -E}_{\zeta\eta}\stackrel{\Omega\ll 1}{\longrightarrow} 2\omega_{0}\zeta\eta\Omega^{2}=\frac{Ev^{2}}{B^{2}_{0}c^{2}}\,\frac{\partial B_{z}}{\partial z}
 \end{equation}
for the $E$-odd frequency shift. This limit is independent of the particle magnetic moment and can be described \cite{PEN01} as a purely geometric Berry phase.

Further, we have verified the direct relationship between the shift $\propto EB$ and that $\propto E^{2}$, which is based on the general identity (\ref{93}) derived in Sec.~\ref{sec:V A.3}. This relationship translates into the identity
\begin{equation}\label{62}
\frac{\langle\delta\omega\rangle^{E\rightarrow -E}_{\zeta\eta}}{\zeta\eta}=4\frac{\langle\delta\omega\rangle_{\eta^{2}}}{\eta^{2}},
\end{equation}
which is clearly seen in Figs.~\ref{fig:two}, \ref{fig:three}, \ref{fig:seven} and \ref{fig:eight}.

\section{Simulations}\label{sec:IV}

We use the Bloch equation \cite{BLO01} in the form of the two coupled ODE's (44) and (45) of Ref.~\cite{PEN01},
\begin{align}\label{79}
&\frac{dJ_{z}}{dt}=J_{xy}\left[\omega_{x}\sin(\Phi+\omega_{0}t)-\omega_{y}\cos(\Phi+\omega_{0}t) \right]\nonumber\\
&\frac{d\Phi}{dt}=-\frac{J_{z}}{J_{xy}}\left[\omega_{x}\cos(\Phi+\omega_{0}t)+\omega_{y}\sin(\Phi+\omega_{0}t) \right],
\end{align}
where $\omega_{x}=\omega_{0}B_{x}/B_{0}$, $\omega_{y}=\omega_{0}B_{y}/B_{0}$, $J_{z}$ ($J_{xy}$) is the vertical (horizontal) spin component and $\Phi$ is the expectation value of its azimuthal angle measured in the system rotating about the $z$ axis at the Larmor frequency $\omega_{0}$. As in \cite{PEN01} we can assume $J_{z}=0$ at the start $t=0$ of the Ramsey measurement period and, since the perturbation is small, that $\Phi$ remains small while $J_{xy}$ remains almost exactly constant. $J_{z}$ oscillates with small amplitude about the equatorial plane.

As in Ref.~\cite{PEN01} we adopt the field model with constant vertical gradient. Including the motional field, the net magnetic field ``seen'' along any chord is, for any cell geometry:
\begin{equation}\label{80}
\frac{B_{x}}{B_{0}}=-\zeta x-\eta\Omega=-\zeta\sigma-\eta\Omega,\,\,\frac{B_{y}}{B_{0}}=-\zeta y=-\zeta\Omega\tau,
\end{equation}
where time $\tau=\omega_{0}t$ is measured from the epicenter $c$ (Fig.~\ref{fig:five}) and the reference length (given by the radius $R$ in case of circular geometry) is set equal to $1$.

Setting $S_{z}(\tau)=J_{z}(\tau)/J_{xy}$ and denoting the cumulative time elapsed since the start of the Ramsey period by $\tau_{\textrm{cum}}$ we obtain from (\ref{79}) and (\ref{80}) the coupled ODE's
\begin{align}\label{81}
&\frac{dS_{z}(\tau)}{d\tau}=\zeta\Omega\tau\cos(\Phi+\tau_{\textrm{cum}}-\Delta\Phi)\nonumber\\
&-(\zeta\sigma+\eta\Omega)\sin(\Phi+\tau_{\textrm{cum}}-\Delta\Phi),\nonumber\\
&\frac{d\Phi(\tau)}{d\tau}=S_{z}(\tau)\big[(\zeta\sigma+\eta\Omega)\cos(\Phi+\tau_{\textrm{cum}}-\Delta\Phi)\nonumber\\
&+\zeta\Omega\tau\sin(\Phi+\tau_{\textrm{cum}}-\Delta\Phi)\big].
\end{align}

In (\ref{81}) we take account of the change $-\Delta\Phi$ of spin angle, as seen in the new reference system after each reflection. $\Delta\Phi$ equals the sum of all phase jumps $\Delta\Phi_{i-1,i}=\alpha_{g,b,i-1}+\alpha_{g,a,i}$ between consecutive chords.

For each chord, we integrate the coupled first-order ODE's (\ref{81}) numerically, with $S_{z}$ and $\Phi$ remaining unchanged at the impact points. To select the $E$-odd phase shift $\Phi_{\zeta\eta}$ this process is carried out separately for positive $\eta$ (corresponding to $\mathbf{E}$ pointing up) and for $\eta$ replaced by $-\eta$ (for $\mathbf{E}$ down). The difference equals $2\Phi_{\zeta\eta}$ and the $E$-odd frequency shift is obtained by division by $\tau_{\textrm{cum}}$. We verified that the asymptotic frequency shifts are independent of initial spin angle $\Phi_{0}$ and of starting position at any point of the cell area, as expected.

In Figs.~\ref{fig:seven} and \ref{fig:eight} we show also the results for the second order shifts $\propto E^{2}$ and $\propto B^{2}$.


\section{Equivalence between the Schr\"odinger equation and the correlation
function approaches}

\label{sec:V}

Note that in this section we mainly use dimensional quantities instead of
dimensionless parameters like $\zeta$, $\eta$ and $\Omega$ in order to
facilitate reference to some previous work. The symbol $\tau$ will represent
the independent variable of the correlation functions and the sign convention
for the Larmor frequency $\omega_{0}$ is different.


\subsection{General proof of equivalence}

\label{sec:V A}

As the novel method of calculation used above differs from the Redfield theory
usually applied in this field we now turn to the relation between the two methods.

In the following we use the direct solutions of the Schr\"{o}dinger equation
for confined spin-$1/2$ particles in arbitrary fields in \cite{STE01} and
\cite{GOL02} to investigate the conditions necessary for the Redfield theory
to be valid.

We consider the solution of the Schr\"odinger equation for arbitrary time
varying fields correct to second order, Eqs.~(34) and (35) of \cite{GOL02} for
the spin-up and spin-down wave functions $g_{r}(t,t_{0})$ and $h_{r}(t,t_{0})$
in the rotating system:
\begin{align}
\label{82} &  g_{r}(t,t_{0})\nonumber\\
&  =1-\frac{1}{4}\int_{t_{0}}^{t}dt^{\prime}\text{e}^{i\omega_{0}t^{\prime}%
}\Sigma^{\ast}(t^{\prime})\int_{t_{0}}^{t^{\prime}}dt^{\prime\prime}%
\text{e}^{-i\omega_{0}t^{\prime\prime}}\Sigma(t^{\prime\prime}),
\end{align}%
\begin{equation}
\label{83}h_{r}(t,t_{0})=-\frac{i}{2}\int_{t_{0}}^{t}dt^{\prime}%
\text{e}^{-i\omega_{0}t^{\prime}}\Sigma(t^{\prime}).
\end{equation}

As above, $\Sigma(t)=\omega_{x}\left(  t\right)  +i\omega_{y}\left(  t\right)
$ is the arbitrary time dependent transverse magnetic field.

Eqs.~(\ref{82}) and (\ref{83}) are reformulations of Eqs.~(17) - (19) of
\cite{STE01} with times $t$, $t^{\prime}$, $t^{\prime\prime}$ no longer
restricted to a single chord but considered a common variable. We have adapted
the terminology of \cite{GOL02} to that used in \cite{STE01} and in the
present article. In \cite{GOL02}, symbols $\alpha_{r}$ and $\beta_{r}$ were
used for the wave functions at any time, not only for a single chord. We
replace them by $g_{r}$ and $h_{r}$ as in (\ref{12}).

\subsubsection{Frequency shifts}\label{sec:V A.1}

The wave function of the spin system, $\psi\left(  t\right)  $ is given by the
matrix (\ref{12}) acting on the initial state. We take that initial state to
be the state corresponding to the spin pointing in the $x$ direction and
evaluate $\left\langle \sigma_{+}\right\rangle _{r}=\left\langle \sigma
_{x}+i\sigma_{y}\right\rangle _{r}$ in the state $\psi\left(  t\right)
.\left\langle \sigma_{+}\right\rangle _{r}$ represents the rotating component
of the spin \ as seen in the rotating coordinate system. The time dependence
of its phase gives the frequency shift and the decay of its magnitude gives
$T_{2.}$ Thus
\begin{align}
\left\langle \sigma_{+}\right\rangle _{r}  &  =\left(  g_{r}^{\ast2}-h_{r}%
^{2}\right) \label{221}\\
\left\langle \sigma_{+}\right\rangle _{r}  &  =1-\frac{1}{2}\int_{0}%
^{t}dt^{\prime}e^{i\omega_{0}t^{\prime}}\Sigma^{\ast}\left(  t^{\prime
}\right)  \int_{0}^{t^{\prime}}dt^{\prime\prime}\Sigma\left(  t^{\prime\prime
}\right)  e^{-i\omega_{0}t^{\prime\prime}}\\
&  =1+z_{2}%
\end{align}
neglecting rapidly varying $h_{r}^{2}$ $\left(  2\omega_{0}\right)  $ term.

We then evaluate the ensemble-averaged frequency shift, using
\begin{align}
\left\langle \delta\phi\right\rangle  &  =\arg\left\langle \sigma
_{+}\right\rangle_{r} =\arg\left(  1+z_{2}\right)  \simeq\operatorname{Im}\left[
z_{2}\right]  \nonumber\\
&  =-\frac{1}{2}\operatorname{Im}\int_{0}^{t}dt^{\prime}e^{i\omega
_{0}t^{\prime}}\Sigma^{\ast}\left(  t^{\prime}\right)  \int_{0}^{t^{\prime}%
}dt^{\prime\prime}\Sigma\left(  t^{\prime\prime}\right)  e^{-i\omega
_{0}t^{\prime\prime}}%
\end{align}%
\begin{align}
\left\langle \delta\omega\right\rangle  &  =\frac{\left\langle \delta
\phi\right\rangle _{T}}{T}\label{84}\\
= &  \frac{1}{2T}\operatorname{Im}\left\langle \int_{0}^{T}dt^{\prime
}\,\text{e}^{i\omega_{0}t^{\prime}}\Sigma^{\ast}(t^{\prime})\,\int
_{0}^{t^{\prime}}dt^{\prime\prime}\,\text{e}^{-i\omega_{0}t^{\prime\prime}%
}\Sigma(t^{\prime\prime})\right\rangle \nonumber
\end{align}
where we have set $t_{0}=0$. $\ T$ is the total elapsed time which, in the
experiments, is given by the long period of free spin precession in the Ramsey
(free induction decay, fid) scheme; thus we are interested in the limit
$T\rightarrow\infty$.

We rewrite (\ref{84}) as%
\begin{align}
\delta\omega &  =\frac{1}{2T}\operatorname{Im}\int_{0}^{T}dt^{\prime}%
\,\int_{0}^{t^{\prime}}dt^{\prime\prime}e^{i\omega_{0}\left(  t^{\prime
}-t^{\prime\prime}\right)  }\left\langle \Sigma^{\ast}(t^{\prime}%
)\,\Sigma(t^{\prime\prime})\right\rangle ,\label{85}\\
\delta\omega &  =\frac{1}{2T}\operatorname{Im}\int_{0}^{T}d\tau\left(
T-\tau\right)  e^{i\omega_{0}\tau}\left\langle \Sigma^{\ast}(0)\,\Sigma
(\tau)\right\rangle , \label{86}%
\end{align}
where the second step is a well known result \cite{SQU01,Papou}.

We have made the usual assumption that the field fluctuations have zero time
average and are stationary, i.e. $\left\langle \Sigma^{\ast}(t^{\prime
})\,\Sigma(t^{\prime\prime})\right\rangle =\left\langle \Sigma^{\ast
}(0)\,\Sigma(\tau=t^{\prime}-t^{\prime\prime})\right\rangle =R(\tau)=R(-\tau
)$. This also applies to the case of specular reflections when we take the
ensemble average over different trajectories (different $\alpha_{g}$'s) and
different starting positions as has been shown in \cite{BAR01}.

In the usual physical situation we have for the observation time $T$: $T\sim
T_{relax}>>\tau_{corr}$ where $T_{relax}$ is the spin relaxation time and
$R\left(  \tau\right)  \rightarrow0$ for $\tau\gtrsim\tau_{corr}$ so that in
this limit
\begin{align}
\delta\omega &  =\frac{1}{2}\operatorname{Im}\int_{0}^{\infty}d\tau
e^{i\omega_{0}\tau}\left\langle \Sigma^{\ast}(0)\,\Sigma(\tau)\right\rangle
\nonumber\label{87}\\
&  =\frac{1}{2}\int_{0}^{\infty}d\tau\cos\omega_{0}\tau\operatorname{Im}%
\left\langle \Sigma^{\ast}(0)\,\Sigma(\tau)\right\rangle \nonumber\\
&  +\frac{1}{2}\int_{0}^{\infty}d\tau\sin\omega_{0}\tau\operatorname{Re}%
\left\langle \Sigma^{\ast}(0)\,\Sigma(\tau)\right\rangle
\end{align}
which is the Redfield result \cite{RED01}.

\subsubsection{Relaxation rate $1/T_{2}$}\label{sec:V A.2}

We now calculate the relaxation rate $1/T_{2}.$

Since $\left|  z_{2}\right|  <<1$%
\begin{align}
\left|  \left\langle \sigma_{+}\right\rangle_{r} \right|  ^{2} &
=1+2\operatorname{Re}z_{2}\\
\left|  \left\langle \sigma_{+}\right\rangle_{r} \right|   &  =1+\operatorname{Re}%
z_{2}\nonumber\\
&  =1-\frac{1}{2}\operatorname{Re}\int_{t_{0}}^{t}dt^{\prime}\int_{t_{0}%
}^{t^{\prime}}dt^{\prime\prime}e^{i\omega_{0}\left(  t^{\prime}-t^{\prime
\prime}\right)  }\left\langle \Sigma^{\ast}\left(  t^{\prime\prime}\right)
\Sigma\left(  t^{\prime}\right)  \right\rangle
\end{align}

We now specialize to the case of a stationary stochastic system where
$\left\langle \Sigma^{\ast}\left(  t^{\prime}\right)  \Sigma\left(
t^{\prime\prime}\right)  \right\rangle $ is a function of $\left(  t^{\prime
}-t^{\prime\prime}\right)  $ only. Consider a square region of the
$t^{\prime\prime},t^{\prime}$ plane between $\left(  t_{0},t_{0}\right)  $,
$\left(  t_{0},t\right)  ,\left(  t,t_{0}\right)  $ and $\left(  t,t\right)
$. (See (\cite{SQU01}) \ for a discussion of this argument). Then the double
integral over the top half $\left(  t^{\prime}>t^{\prime\prime}\right)  $ is
seen to be the complex conjugate of the integral over the bottom half $\left(
t^{\prime}<t^{\prime\prime}\right)  $, so the last term is given by the
integral over the entire square.

As a result of this we have\bigskip\ (again putting $t^{\prime}-t^{\prime
\prime}=\tau$)%
\begin{align}
\left|\left\langle \sigma_{+}\right\rangle_{r}\right|   &  =1-\frac{1}{2}\operatorname{Re}\int_{0}%
^{t}dt^{\prime}\int_{0}^{t}dt^{\prime\prime}e^{i\omega_{0}\left(  t^{\prime
}-t^{\prime\prime}\right)  }\left\langle \Sigma^{\ast}\left(  t^{\prime
}\right)  \Sigma\left(  t^{\prime\prime}\right)  \right\rangle \nonumber\\
&  =1-\frac{1}{2}\operatorname{Re}\int_{0}^{t}d\tau\left(  t-\left|
\tau\right|  \right)  e^{i\omega_{0}\tau}\left\langle \Sigma^{\ast}\left(
t^{\prime}\right)  \Sigma\left(  t^{\prime}-\tau\right)  \right\rangle
\nonumber\\
&  =1-\frac{1}{2}t\operatorname{Re}\int_{0}^{t}d\tau e^{i\omega_{0}\tau
}\left\langle \Sigma^{\ast}\left(  t^{\prime}\right)  \Sigma\left(  t^{\prime
}-\tau\right)  \right\rangle \label{oooa}\\
\left|\left\langle \sigma_{+}\right\rangle_{r}\right|   &  =1-\frac{t}{T_{2}}\label{00a}%
\end{align}
where the result is valid for times $t>\tau_{corr}$. Substituting
\begin{equation}
\Sigma\left(  t\right)  =\omega_{x}\left(  t\right)  +i\omega_{y}\left(
t\right)  \label{bg1}%
\end{equation}
we obtain the usual Redfield result:%
\begin{align}
\frac{1}{T_{2}} &  =\frac{1}{2}\operatorname{Re}\int_{0}^{\infty}d\tau
e^{i\omega_{0}\tau}\left\langle \Sigma^{\ast}\left(  t^{\prime}\right)
\Sigma\left(  t^{\prime}-\tau\right)  \right\rangle \\
&  =\frac{1}{2}\int_{0}^{\infty}d\tau\left(
\begin{array}
[c]{c}%
\cos\omega_{0}\tau\left\langle
\begin{array}
[c]{c}%
\omega_{x}\left(  t\right)  \omega_{x}\left(  t-\tau\right)  \\
+\omega_{y}\left(  t\right)  \omega_{y}\left(  t-\tau\right)
\end{array}
\right\rangle \\
+\sin\omega_{0}\tau\left\langle
\begin{array}
[c]{c}%
\omega_{y}\left(  t\right)  \omega_{x}\left(  t-\tau\right)  \\
-\omega_{x}\left(  t\right)  \omega_{y}\left(  t-\tau\right)
\end{array}
\right\rangle
\end{array}
\right)  \label{222}%
\end{align}
The second term is absent in the usual treatments as it is normally assumed
that the cross correlation between the components of the fluctuating field
vanishes but this does not hold in the presence of the $\mathbf{E}%
\times\mathbf{v}$ field.

Thus we have shown that the Redfield expressions for the frequency shift and
$T_{2}$ relaxation rate, see also \cite{MCG01}, follow directly from the
second order solution to the Schr\"{o}dinger equation for arbitrary time
dependent field with zero time average.

To calculate the longitudinal relaxation time, $T_{1},$ we calculate
$\psi\left(  t\right)  $ starting in the spin up state and and then evaluate
\ $\left\langle \sigma_{z}\right\rangle_{r} =\left|  g_{r}\right|  ^{2}-\left|
h_{r}\right|  ^{2}$ with the well known result $1/T_{1}=2/T_{2}^{\prime}$
where $1/T_{2}^{\prime}$ is the transverse relaxation rate in the absence of
fluctuations in $B_{z},$ \cite{GOL02}.

\subsubsection{Expressions for the frequency shifts}

Using (\ref{bg1}) we find%
\begin{equation}
\Sigma=\left(  \gamma B_{x}-bv_{y}\right)  +i\left(  \gamma B_{y}%
+bv_{x}\right)
\end{equation}
where, as above, we are considering a magnetic field $\mathbf{B}$ in the
presence of a motional magnetic field ($\mathbf{E}\times\mathbf{v})/c^{2}$
produced by an electric field in the $z$ direction parallel to $\mathbf{B}$
(we use $b=\gamma E/c^{2}$ and $\gamma=\omega_{0}/B_{0}$ is the gyromagnetic
ratio of the particle)

\label{sec:V A.3}%

\begin{align}
&  \operatorname{Re}\left\langle \Sigma^{\ast}(0)\,\Sigma(\tau)\right\rangle
=\big\langle\gamma^{2}\left(  B_{x}\left(  0\right)  B_{x}\left(  \tau\right)
+B_{y}\left(  0\right)  B_{y}\left(  \tau\right)  \right)  \nonumber\label{88}%
\\
&  +b^{2}\left(  v_{x}\left(  0\right)  v_{x}\left(  \tau\right)
+v_{y}\left(  0\right)  v_{y}\left(  \tau\right)  \right)  \big\rangle
,\nonumber\\
&  \operatorname{Im}\left\langle \Sigma^{\ast}(0)\,\Sigma(\tau)\right\rangle
=-b\gamma\big\langle v_{x}\left(  0\right)  B_{x}\left(  \tau\right)
-B_{x}\left(  0\right)  v_{x}\left(  \tau\right) \nonumber\\
&  +v_{y}\left(  0\right)  B_{y}\left(  \tau\right)  -B_{y}\left(  0\right)
v_{y}^{{}}\left(  \tau\right)  \big\rangle\nonumber\\
&  =2b\gamma\left\langle B_{x}\left(  0\right)  v_{x}^{{}}\left(
\tau\right)  +B_{y}\left(  0\right)  v_{y}^{{}}\left(  \tau\right)
\right\rangle
\end{align}
where we have assumed that the motions in the $x$ and $y$ directions are
uncorrelated. Compare \cite{PIG01}, equations (8) - (10). Specializing to the
case of a linear magnetic gradient, $\partial B_{x}/\partial x=\partial
B_{y}/\partial y=G$, we have%
\begin{align}
&  \operatorname{Re}\big\langle\Sigma^{\ast}(0)\,\Sigma(\tau)\big
\rangle=\gamma^{2}G^{2}\big\langle x\left(  0\right)  x\left(  \tau\right)
+y\left(  0\right)  y\left(  \tau\right)  \big\rangle\nonumber\label{89}\\
&  +\left(  \gamma E/c^{2}\right)  ^{2}\big\langle v_{x}\left(  0\right)
v_{x}\left(  \tau\right)  +v_{y}\left(  0\right)  v_{y}\left(  \tau\right)
\big\rangle,\nonumber\\
&  \operatorname{Im}\left\langle \Sigma^{\ast}(0)\,\Sigma(\tau)\right\rangle
\nonumber\\
&  =2\gamma^{2}(GE/c^{2})\big\langle x\left(  0\right)  v_{x}^{{}}\left(
\tau\right)  +y\left(  0\right)  v_{y}^{{}}\left(  \tau\right)  \big\rangle.
\end{align}

Thus according to equations (\ref{87}) and (\ref{89}), see also \cite{LAM01},
the frequency shift $\sim E^{2}$ is given by
\begin{align}
\label{90} &  \delta\omega_{E^{2}}=\left(  \gamma E/c^{2}\right)
^{2}\operatorname{Im}\left[  \bar{S}_{vv}\left(  \omega_{0}\right)  \right]
,\\
&  \bar{S}_{vv}\left(  \omega_{0}\right)  =\int_{0}^{\infty}d\tau e^{i\omega_{0}\tau}S_{vv}(\tau)\nonumber\\
&\textrm{with}\,\,S_{vv}(\tau)=\left\langle v_{x}\left(  0\right)  v_{x}\left(
\tau\right)  +v_{y}\left(  0\right)  v_{y}\left(  \tau\right)  \right\rangle
,\nonumber
\end{align}
while the linear in $E$ shift is given by
\begin{align}
&  \delta\omega_{EB}=
2\gamma^{2}(GE/c^{2})\operatorname{Re}\bar{S}_{xv}(\omega_{0})\nonumber\\
&\textrm{with}\,\,\bar{S}_{xv}(\omega_{0})=\int_{0}^{\infty}d\tau e^{i\omega_{0}\tau}S_{xv},\label{91}\\
&S_{xv}\left(  \tau\right)=\big\langle x\left(  0\right)  v_{x}^{{}}\left(
\tau\right)  +y\left(  0\right)  v_{y}^{{}}\left(  \tau\right)  \big\rangle= \int_{0}^{\tau}S_{vv}\left(  \mu\right)
d\mu\nonumber  ;\\
&\textrm{thus}\,\,\bar{S}_{xv}\left(  \omega_{0}\right)=\int_{0}^{\infty}d\tau e^{i\omega_{0}\tau}\int_{0}^{\tau} S_{vv}\left(
\mu\right)  d\mu \nonumber\\
&=\frac{i}{\omega_{0}}\int_{0}^{\infty}d\tau e^{i\omega_{0}\tau}S_{vv}\left(
\tau\right) = \frac{i}{\omega_{0}}\bar{S}_{vv}\left(  \omega_{0}\right)\label{91a}\\
& \textrm{and similarly,}\,\,\bar{S}_{xx}\left(  \omega_{0}\right)=-\frac{i}{\omega_{0}}\bar{S}_{xv}\left(  \omega_{0}\right)=\frac{1}{\omega^{2}_{0}}\bar{S}_{vv}\left(  \omega_{0}\right).\label{91b}
\end{align}
Substituting (\ref{91a}) in (\ref{91}) we obtain
\begin{align}
&\delta\omega_{EB}=-2\frac{\gamma^{2}GE}{\omega
_{0}c^{2}}\operatorname{Im}\left[  \bar{S}_{vv}\left(  \omega_{0}\right)
\right]  \label{92}%
\end{align}
and comparing (\ref{92}) with (\ref{90}) we find
\begin{equation}
\label{93}\frac{\delta\omega_{EB}}{\delta\omega_{E^{2}}}=-2\frac{Gc^{2}%
}{E\omega_{0}},
\end{equation}
i.~e.~the frequency dependence of the two effects differs only by the factor
$\omega_{0}$. This was noted by Pendlebury \emph{et al.} \cite{PEN01} for the
case of a circular cylinder with specular walls but our result is universal.
It holds for all geometries, field configurations and collisional regimes and is clearly shown in Figs.~\ref{fig:two}, \ref{fig:three}, \ref{fig:seven} and \ref{fig:eight}.
Using (\ref{91a}) we can also show that the result given in Eq.~(64) of \cite{JEE01} does indeed agree with previously published results contrary to the doubt expressed in \cite{JEE01}.

\subsection{Correlation functions for a circular cylinder with diffuse wall collisions}

\label{sec:V B}

We calculate the velocity correlation function for a circular vessel with
diffuse reflecting walls as follows.

We consider a trajectory that starts on a chord with $\alpha_{g,0}$
(Fig.~\ref{fig:one}) at a distance $x$ from the end of the chord. The particle
reaches the end of the chord at time $t_{n=0}$. It then travels on a chord
$\alpha_{g,1}$ until a time $t_{1}$. Between $t_{n}$ and $t_{n+1}$ the chord
angle is $\alpha_{g,n+1.}$ At each wall collision at $t_{n}$ the velocity
vector rotates through an angle $\delta\theta=\alpha_{g,n+1}+\alpha_{g,n}.$
For negative $\rho_{n+1}$ (motion in the positive CCW direction),
$\delta\theta=\pi+\rho_{n}+\rho_{n+1}$ for both positive and negative
$\rho_{n}$ while for positive $\rho_{n+1}$ (motion in the CW direction) we
have $\delta\theta=-\pi+\rho_{n}+\rho_{n+1}$. Since $\pi=-\pi$ so long as
angles are concerned all that matters is whether we have an odd or even number
of $\pi^{\prime}s$. This will result in a varying sign when we take the sine
or cosine of the angle. We have that between $t_{n}$ and $t_{n+1}$ the
velocity makes an angle $\theta_{n}$ with its original direction with
\begin{equation}
\label{94}\cos\theta_{n}=\left(  -1\right)  ^{n+1}\cos\left(  \rho_{0}%
+\rho_{n+1}+2\sum_{k=1}^{n}\rho_{k}\right)  .
\end{equation}
The wall collisions occur at times%
\begin{equation}
\label{95}t_{n}=\frac{2R}{v}\left(  \sum_{k=1}^{n}\cos\rho_{k}\right)  +t_{0}%
\end{equation}
with $t_{0}=x/v.$

So the correlation function, after $\tau=t_{0}$, consists of a series of
pulses between times $t_{n}$ and $t_{n+1}$ with heights (normalized to $v^{2}
$) $\cos\theta_{n}.$

The Fourier transform $\bar{S}_{vv}\left(  \omega\right)  $ (\ref{90}) will be
a sum of terms each corresponding to one of the rectangular pulses. Each term
is given by the Fourier transform of a rectangular pulse,
\begin{align}
\label{96}S_{n}\left(  \omega\right)   &  =2\cos\theta_{n}\frac{\sin
\frac{\omega(\delta t)_{n}}{2}}{\omega}e^{-i\omega\left(  \frac{t_{n+1}+t_{n}%
}{2}\right)  }\nonumber\\
&  =2\cos\theta_{n}\frac{\sin\delta_{n+1}}{\omega}e^{-i\omega\left(
\frac{t_{n+1}+t_{n}}{2}\right)  }\\
&  =2e^{-i\omega x/v}\cos\theta_{n}\frac{\sin\delta_{n+1}}{\omega}e^{-i\left(
\delta_{n+1}+ 2\sum_{k=1}^{n}\delta_{k}\right)  },\nonumber
\end{align}
where $\delta_{n}=\frac{\omega R}{v}\cos\rho_{n}$ and
\begin{equation}
\label{97}(\delta t)_{n}=t_{n+1}-t_{n}=\frac{2R}{v}\left(  \cos\rho
_{n+1}\right)  .
\end{equation}
We have used, from (\ref{95}),
\begin{equation}
\label{98}\frac{t_{n}+t_{n+1}}{2}=\frac{R}{v}\left(  \cos\rho_{n+1}%
+2\sum_{k=1}^{n}\cos\rho_{k}\right)  +t_{0}.
\end{equation}

We have to average the result over all possible starting positions and
directions of motion. That means we average over all possible initial chord
angles, $\alpha_{g,0},$ and all positions along the chord given by the
distance $x$ from the end. $x$ varies between $x=0$ and $x=2 R\cos\rho.$
Averaging (\ref{96}) over $x$ gives
\begin{align}
\label{99} &  \frac{1}{2R\cos\rho_{0}}\int_{0}^{2R\cos\rho_{0}}e^{-i\omega
x/v}dx\nonumber\\
&  =\frac{v}{\omega}\frac{1}{R\cos\rho_{0}}\sin\left(  R\left(  \cos\rho
_{0}\right)  \frac{\omega}{v}\right)  e^{-i\left(  R\left(  \cos\rho
_{0}\right)  \frac{\omega}{v}\right)  }\nonumber\\
&  =\frac{\sin\delta_{0}}{\delta_{0}}e^{-i\delta_{0}},\\
\bar{S}_{n}\left(  \omega\right)   &  =2\frac{\sin\delta_{0}}{\delta_{0}%
}e^{-i\delta_{0}}\cos\theta_{n}\frac{\sin\delta_{n+1}}{\omega}e^{-i\left(
\delta_{n+1}+2\sum_{k=1}^{n}\delta_{k}\right)  }.\nonumber
\end{align}

We write $\cos\theta_{n}=\left(  e^{i\theta_{n}}+e^{-i\theta_{n}}\right)  /2$
and consider only the first term. We will then add in the second term at the end.

Summing $\bar{S}_{n}\left(  \omega\right)  $ from (\ref{99}) over all chords
except the first (labeled $00$) we have, using (\ref{94}),
\begin{align}
\label{100} &  -\frac{2}{\omega}\frac{\sin\delta_{0}}{\delta_{0}}e^{i\left(
\rho_{0}-\delta_{0}\right)  }\\
&  \times\,\sum_{n=0}^{N}\left(  -1\right)  ^{n}\sin\delta_{n+1}e^{i\left(
\rho_{n+1}-\delta_{n+1}\right)  }e^{2i\sum_{k=1}^{n}\left(  \rho_{k}%
-\delta_{k}\right)  }.\nonumber
\end{align}
For $j=1\,...\,n$ the angles $\rho_{j}$ are independent of $\rho_{0}$,
$\rho_{n+1}$ and of each other, so the average of (\ref{100}) can be written%
\begin{align}
&  =-\frac{2}{\omega}\left\langle \frac{\sin\delta_{0}}{\delta_{0}}e^{i\left(
\rho_{0}-\delta_{0}\right)  }\right\rangle \left\langle \sin\delta
_{n+1}e^{i\left(  \rho_{n+1}-\delta_{n+1}\right)  }\right\rangle \nonumber\\
&  \times\,\sum_{n=0}^{N}\left(  -1\right)  ^{n}\left\langle e^{2i\left(
\rho_{j}-\delta_{j}\right)  }\right\rangle ^{n}\nonumber\\
&  =-\frac{2}{\omega}\left\langle \frac{\sin\delta_{0}}{\delta_{0}}e^{i\left(
\rho_{0}-\delta_{0}\right)  }\right\rangle \left\langle \sin\delta
_{n+1}e^{i\left(  \rho_{n+1}-\delta_{n+1}\right)  }\right\rangle \nonumber\\
&  \times\,\frac{1}{1+\left\langle e^{2i\left(  \rho_{j}-\delta_{j}\right)
}\right\rangle }.\nonumber
\end{align}
To this we add the contribution of the second term, $\text{e}^{-i\theta_{n}}$,
in the expansion of $\cos\theta_{n}$ and the contribution of the first chord
which adds a term equal to $v^{2}$ for a time from $t=0\,$ to $t=t_{0}=x/v$
and will have a spectrum%
\begin{equation}
\label{101}S_{00}=2\frac{\sin\left(  \frac{\omega x}{2v}\right)  }{\omega
}e^{-i\omega\left(  \frac{x}{2v}\right)  }.
\end{equation}
Its average over $x$ is
\begin{align}
\label{102} &  \bar{S}_{00}(\omega)=\frac{1}{\omega R\cos\rho_{0}}\int
_{0}^{2R\cos\rho_{0}}\sin\left(  \frac{\omega x}{2v}\right)  e^{-i\omega
\left(  \frac{x}{2v}\right)  }dx\nonumber\\
&  =\frac{1}{\omega\delta_{0}}\left(  \frac{1}{2}-\frac{1}{2}e^{-2i\delta_{0}%
}-i\delta_{0}\right)
\end{align}
and the complete expression for the spectrum is
\begin{align}
\label{103} &  \bar{S}_{vv}\left(  \omega\right)  =-\frac{2}{\omega
}\left\langle \frac{\sin\delta_{0}}{\delta_{0}}e^{i\left(  \rho_{0}-\delta
_{0}\right)  }\right\rangle \nonumber\\
&  \times\,\left\langle \sin\delta_{n+1}e^{i\left(  \rho_{n+1}-\delta
_{n+1}\right)  }\right\rangle \frac{1}{1+\left\langle e^{2i\left(  \rho
_{j}-\delta_{j}\right)  }\right\rangle }\nonumber\\
&  +\left[  \left(  \rho_{j}-\delta_{j}\right)  \rightarrow\left(  \rho
_{j}-\pi-\delta_{j}\right)  \right] \nonumber\\
&  +\left\langle \frac{1}{\omega\delta_{0}}\left(  \frac{1}{2}-\frac{1}%
{2}e^{-2i\delta_{0}}-i\delta_{0}\right)  \right\rangle .
\end{align}
The average has to be taken over all possible starting positions and
directions of velocity $\left(  x,\rho_{0}\right)  .$ Since the number of
particles located on a given chord is proportional to the length of the chord
$\left(  l_{j}=2R\cos\rho_{j}\right)  $ we use averaging method 1 for the
first chord. However because following each wall collision the angles
$\rho_{j}$ are distributed by a cosine law (Lambert's law in 2 dimensions) we
use method 2 for all other chords. Then using Eq. (\ref{90}) we can write
\begin{align}
\label{104} &  \frac{\delta\omega_{E^{2}}}{\left(  \gamma v E/c^{2}\right)
^{2}}=\operatorname{Im}\left[  \bar{S}_{vv}\left(  \omega_{0}\right)  \right]
\nonumber\\
&  =\operatorname{Im}\left\{
\begin{array}
[c]{c}%
-\frac{2}{\omega_{0}}\frac{\left(  \left\langle \sin\delta_{0}e^{i\left(
\alpha_{g,0}-\delta_{0}\right)  }\right\rangle _{2}\right)  ^{2}}{\left\langle
\delta_{0}\right\rangle _{2}}\frac{1}{1-\left\langle e^{2i\left(  \alpha
_{g,j}-\delta_{j}\right)  }\right\rangle _{2}}\\
+\left[  \left(  \alpha_{g,j}-\delta_{j}\right)  \rightarrow\left(
-\alpha_{g,j}+\pi-\delta_{j}\right)  \right]
\end{array}
\right\} \nonumber\\
&  +\frac{1}{\omega_{0}\left\langle \delta_{0}\right\rangle _{2}}\left\langle
\left(  \frac{1}{2}\left(  \sin2\delta_{0}\right)  -\delta_{0}\right)
\right\rangle _{2}%
\end{align}
in agreement with the Eq.~(\ref{37}), taking into account a different sign
convention for $\omega_{0}$. We have included the normalization factor $v^{2}
$ in (\ref{104}) and, on the last term, used (\ref{33}) and again (\ref{90}).

\subsection{Correlation functions and frequency shifts for a rectangular cell
with specular walls}

\label{sec:V C}

Substituting expression (\ref{89}) for $\operatorname{Im}\left\langle
\Sigma^{\ast}(0)\,\Sigma(\tau)\right\rangle $ in equation (\ref{87}), noting
that
\begin{equation}
\label{105}\left\langle x\left(  0\right)  v_{x}^{{}}\left(  \tau\right)
\right\rangle =\frac{\partial}{\partial\tau}\left\langle x\left(  0\right)
x\left(  \tau\right)  \right\rangle
\end{equation}
and integrating by parts we find for the linear in $E$ frequency shift
\cite{SwankPhD}:
\begin{align}
\label{106}\delta\omega_{EB}  &  =-\gamma^{2}G\frac{E}{c^{2}}\Big[\frac{1}%
{12}\left(  L_{x}^{2}+L_{y}^{2}\right) \\
&  -\omega_{0}\int_{0}^{\infty}d\tau\sin\omega_{0}\tau\left\langle x\left(
0\right)  x\left(  \tau\right)  +y\left(  0\right)  y\left(  \tau\right)
\right\rangle \Big],\nonumber
\end{align}
a result first obtained in \cite{PIG01}. Swank $\textit{et al.}$ \cite{SWA01}
(Eq.~(17)), using the form of transport propagator derived by Masoliver
\emph{et al.}~\cite{Mas}, have given an analytic result for the Fourier
transform, $S_{xx}\left(  \omega\right)  ,$ of the correlation function
$\left\langle x\left(  0\right)  x\left(  \tau\right)  \right\rangle $ in the
case of a rectangular cell with specularly reflecting walls and gas
collisions, which together with (\ref{106}) yields:
\begin{align}
\label{107} &  \delta\omega_{EB}=-\gamma^{2}G\frac{E}{c^{2}}\Big[\frac{1}%
{12}\left(  L_{x}^{2}+L_{y}^{2}\right)  +\\
&  \operatorname{Im}\left\{  \!\!
\begin{array}
[c]{c}%
\frac{4}{\pi^{4}}L_{x}^{2}\sum_{n_{x}=\text{odd}} \,n^{-4}_{x}\frac{\omega
_{0}\tau_{c}}{\left[  \left(  1+i\omega_{0}\tau_{c}\right)  ^{2}+\left(
n_{x}\pi l_{x}\right)  ^{2}\right]  ^{1/2}-1}\\
+\left(  x\rightarrow y\right)
\end{array}
\!\!\right\}  \Big],\nonumber
\end{align}
where $l_{x,y}=v\tau_{c}/L_{x,y}=\lambda_{c}/L_{x,y},$ $L_{x,y}$ are the
lengths of the sides, $v$ is the velocity, $\tau_{c}$ is the average time
between gas collisions and $\lambda_{c}$ is the mean free path. This is valid
for all values of the mean free path, so taking the ballistic limit
$l_{x,y}\gg1$ we obtain the result plotted in Fig. \ref{fig:eight} with
reference to Swank $\textit{et al.}$ \cite{SWA01}. More details can be found
in \cite{GOLetal}.


\section{Summary and conclusions}

\label{sec:VI}

Solving the Schr\"odinger equation directly we have analyzed the frequency
shifts in searches for an electric dipole moment of the neutron using
ultracold neutrons and co-magnetometer atoms confined in measurement cells.
The cells are assumed to have a perfectly flat top and bottom and can have
rough cylinder walls (Sec.~\ref{sec:II}) and a cross section of generic shape
(Sec.~\ref{sec:III A}). For perfectly diffuse 2D roughness scattering on the
walls and circular cell cross section we derived expression (\ref{36}) which
is valid for any in-plane particle velocity $v$. Its generalization to
arbitrary cell geometry in the form of Eq.~(\ref{52}) is an approximation.

We have specifically investigated the geometry of a general rectangle in
Sec.~\ref{sec:III B}) and compared all results with the simulations described
in Sec.~\ref{sec:IV}. So far, general results valid in a broad range of
particle velocity had been reported only for circular and rectangular geometry
and specular reflection \cite{PEN01,BAR01,SWA01,STE01}.

In Sec.~\ref{sec:II D} we showed that a Larmor field slightly tilted away from
the vertical symmetry axis of the cell, due to misalignment inevitable in the
experiments, can be taken into account as in Ref.~\cite{LAM01}: The shifts
remain unchanged but refer to the Larmor frequency $\omega^{\prime}_{0}$ given
by the field magnitude, rather than to $\omega_{0} $.

The adiabatic limit ($\Omega\ll1$) of the $E$-odd shift agrees with the purely
geometric Berry phase while those for the second-order frequency shifts,
$\propto E^{2}$ and $\propto B^{2}$, agree with the general results of
\cite{GUI01}. This consistency and, generally, the equivalence between
previous approaches and our analysis are elaborated in Sec.~\ref{sec:V}. We
show by direct calculation the agreement of the results from the Schr\"odinger
equation with the Redfield theory for the cases of a circular cell with
diffuse reflecting walls and of a rectangular cell with specular walls.

These comparisons serve to highlight the conditions of validity of the
Redfield theory which makes use of a number of assumptions \cite{ABR01} and
has resulted in a considerable literature concerning its range of validity,
e.~g.~\cite{NIC01}. Our results can be applied to cases where the Redfield
result no longer holds, such as non-stochastic or partially stochastic
systems. An example of the latter is the inevitable situation where the static
fields $\mathbf{E}$ and $\mathbf{B}_{0}$ are not exactly aligned or where, for
non-spherical cells, $\mathbf{B}_{0}$ is not exactly aligned with the symmetry axis.

Our method also applies to cases where the observation time is on the order of
or shorter than the correlation time, and thus we can describe the transient
spin dynamics, i.e.~the gradual development of the shift with increasing time
subsequent to the start of the free precession in the Ramsey scheme. This
information is more detailed than the transverse relaxation time $T_{2}$
derived from the Redfield theory or directly from the Schr\"odinger equation
\cite{GOL02} or the Bloch equation \cite{JEE01}. Similarly, studying the
vertical spin oscillations as in \cite{STE01} can provide information not
contained in $T_{1}.$ Our analysis in the present article does not include
gravitational depolarization \cite{HAR03,AFA02} due to the effect of gravity
on the UCN spectra in a specific apparatus.

%
\acknowledgments
%

We are grateful to M. Bales, E. Gutsmiedl, P. Fierlinger, P.~G.~Harris and G.~Zsigmond for very helpful discussions.

\appendix

\section{Analysis of rectangular cell geometry}\label{sec:B}

To cover all possible trajectories for the rectangular geometry shown in Fig.~\ref{fig:six} it suffices to analyze the four groups of chords listed in Table \ref{table:two}. These start from side IV (or equivalently, from II) in any direction ($-\pi/2\,<\rho<\pi/2$). The paths of group $1$ ($4$) end on side I (III). We have divided those ending on side II into groups $2$ and $3$. They are separated by the angle $\rho_{\textrm{CM}}(x,w)=\arctan\left[x/(1-w)\right]$ for the path passing through CM. Angle $\rho_{1}(x,w)=-\arctan \left[(1+w-x)/(2(1-w)\right]$ is directed towards the upper right corner. For trajectories originating from sides I or III we replace $w$ by $-w$. 
   
\begin{table}[htb]
\caption{$\rho$-ranges for the four path groups for a rectangle}
\centering
\begin{tabular}{c c c}
\hline\hline
Group $j$ & $\rho$-range  \\[0.5ex]
\hline
1 & $-\frac{\pi}{2}<\rho\leq \rho_{1}(x,w)$ \\
2 & $\rho_{1}(x,w)<\rho<\rho_{\textrm{CM}}(x,w)$ \\
3 & $\rho_{\textrm{CM}}(x,w)<\rho<-\rho_{1}(-x,w)$\\
4 & $-\rho_{1}(-x,w)\leq\rho<\frac{\pi}{2}$\\ [1ex]
\hline
\end{tabular}
\label{table:two}
\end{table}

The characteristics for each path are as follows:
\begin{align}
&L_{1}(x,\rho,w)=-(1+w-x)/\sin\rho\label{B1}\\ 
&L_{j}(x,\rho,w)=2(1-w)/\cos\rho,\,\,j=2,3\label{B2}\\ 
&L_{4}(x,\rho,w)=(1+w+x)/\sin\rho.\label{B3}
\end{align}

For $j=1,...,4$ we have
\begin{align}\label{B4}
&\sigma_{j}(x,\rho,w)=\left(x\cos\rho-(1-w)\sin\rho\right)\,\textrm{sign}(\Omega)\\
&\cos(2\alpha_{g,j})=-\frac{V_{j}}{\sqrt{V^{2}_{j}+W^{2}_{j}}},\,\,\sin(2\alpha_{g,j})=\frac{W_{j}}{\sqrt{V^{2}_{j}+W^{2}_{j}}},\nonumber
\end{align}
\begin{align}\label{B5}
V_{j}(x,\rho,w)&=[x\sin\rho+(1-w)\cos\rho]L_{j}\nonumber\\
&-x^{2}-(1-w)^{2},\nonumber\\
W_{j}(x,\rho,w)&=[x\cos\rho-(1-w)\sin\rho]L_{j},
\end{align}
\begin{align}\label{B6}
&\alpha_{g,j}(x,\rho,w,\Omega)=\frac{1}{2}\textrm{sign}(\Omega)\arccos(\cos 2\alpha_{g,j}),\,(\rho<\rho_{\textrm{CM}})\nonumber\\
&=\frac{1}{2}\textrm{sign}(\Omega)\left[2\pi-\arccos(\cos 2\alpha_{g,j})\right],\,(\rho >\rho_{\textrm{CM}});
\end{align}
\begin{align}\label{B7}
&\alpha_{g,a,j}(x,\rho,w,\Omega)=\big(\frac{\pi}{2}+\rho -\rho_{\textrm{CM}}\big)\textrm{sign}(\Omega),\nonumber\\
&\alpha_{g,b,j}=2\alpha_{g,j}-\alpha_{g,a,j},\nonumber\\
&\delta_{j}(x,\rho,w,\Omega)=L_{j}(x,\rho,w)/(2\Omega),\nonumber\\
&\delta_{a,j}(x,\rho,w,\Omega)=\frac{1}{\Omega}\left[(1-w)\cos\rho +x\sin\rho \right],\nonumber\\
&\delta_{b,j}=2\delta_{j}-\delta_{a,j},\nonumber\\
&\delta_{ba,j}=\delta_{j}-\delta_{a,j}.
\end{align}
The signs appropriate to the different ranges of $\rho$ and to backward vs.~forward motion are shown explicitly. For the arccosine function we use the range from $0$ to $\pi$.

Adapting Eqs.~(\ref{53}) to (\ref{56}) to rectangular geometry we average the frequency shift (\ref{52}) over the four groups as
\begin{align}\label{B8}
&\big\langle(...)\big\rangle_{2,\textrm{even}}=\frac{1}{4 N_{2}}\Big\{\sum_{j=1}^{4}\,\int_{-1-w}^{1+w}dx\!\!\int_{\rho-\textrm{range}}\!\!d\rho\,\cos\rho\nonumber\\
&\times\,\left[(...)_{\textrm{fw}}+(...)_{\textrm{bw}}\right]+\left(w\rightarrow -w \right)\Big\},
\end{align} 
with normalization constant $N_{2}=4$, and the product term in (\ref{52}) in the form
\begin{align}\label{B9}
&\langle F G\rangle_{3}=\frac{1}{2 N_{3}}\Big\{\sum_{j=1}^{4}\,\,
\int_{-1-w}^{1+w}\,dx\,\int_{\rho-\textrm{range}}\,d\rho\,\cos\rho\nonumber\\
&\times\,\left(\left[ F\right]_{\textrm{fw}}\left[ G\right]_{\textrm{fw}}+\left[ F\right]_{\textrm{bw}}\left[ G\right]_{\textrm{bw}}\right)+\left(w\rightarrow -w\right)\Big\},
\end{align}
with $N_{3}=8$. Some averages can be determined analytically, e.~g.~$\langle\delta \rangle_{2}=\pi(1-w^{2})/(4\Omega)$, but most integrations had to be performed numerically.

\end{document}